# Annually Modulated Pacific Subtropical Highs Pace the ENSO Cycle: A Hierarchical Framework from Earth Rotation to Atmosphere–Ocean Dynamics

Yafei Wang*

Correspondence and requests for materials should be addressed to:

*Dr. Yafei Wang

Chinese Academy of Meteorological Sciences

Email: yfwang@cma.gov.cn

## Abstract

The El Niño–Southern Oscillation drives Earth's dominant interannual climate signal, yet what initiates an event and what ends it remain incompletely understood. Here we show that anomalies of the eastern Pacific subtropical high (EPSH) belts, marched through the year by the annual cycle, precede and govern the equatorial zonal wind near the dateline, which sets subsequent sea surface temperature evolution. The Northern Hemisphere ridge triggers events in boreal spring, the Southern Hemisphere ridge sustains them in austral spring, and their synchrony decides ENSO diversity. Preceding the ridges lies a sub-annual oscillation in Earth's rotation rate (length of day, 0.5 to 1 yr band) that redistributes atmospheric mass poleward three times once its phase aligns with the annual cycle, in August, in December, and in the following May, each displacement carrying the response further east and poleward until the significant centers overlie the ridges themselves. Crucially, this band is deaf to ENSO: its cross-correlation with NINO3 never exceeds 0.05 at any lag, whereas the classical ENSO-driven rotation response resides in the 2 to 10 yr band at 0.40. The band therefore reads forward, and partial correlation identifies the EPSH as the conduit: the direct path from rotation to ocean vanishes once the ridges are held fixed. Complete by the July preceding the event year, the index fixes the phase of the coming event fourteen to seventeen months ahead, though not its amplitude. At maturity the event reorganizes the Hadley and Walker circulations into the Mirror-SIGMA pattern, disrupting the upstream chain and closing the cycle.

## Significance Statement

El Niño and La Niña events reshape global weather, agriculture, and economies, yet forecasters still struggle to predict them more than a few months ahead. This study identifies a chain of influence that runs from Earth's rotation rate through the subtropical Pacific highs to the equatorial ocean. The annual cycle acts as a metronome, driving the seasonal migration of the subtropical highs and locking ENSO events to specific times of year. Superimposed on this rhythm, a sub-annual signal in Earth's Length of Day preconditions the Pacific subtropical highs and fixes the phase of the coming event fourteen to seventeen months before its peak, an ENSO precursor independent of ocean-based indicators and available well before the spring predictability barrier.

## 1. Introduction

### *a. The unresolved puzzles of ENSO: a review of classical theories*

The El Niño–Southern Oscillation (ENSO) is the dominant mode of interannual variability in the global climate system (Timmermann et al. 2018), with profound impacts on natural environments and human societies. These include widespread droughts and floods (Ropelewski and Halpert 1987; Cai et al. 2015), disruptions to agriculture and water resources, and associations with extreme weather that ripple across continents. ENSO's complexity stems from the intricate interplay of seasonal variations, trade winds, oceanic waves, and basin geometry, making its dynamical characterization one of the enduring challenges in climate science (McPhaden et al. 2006).

For over a century, scientists have sought to uncover these dynamical mechanisms, producing a series of classical theories that form the cornerstone of our current understanding. A pivotal breakthrough occurred in the 1960s when Bjerknes (1966, 1969) identified El Niño as a manifestation of coupled atmosphere–ocean interaction. This feedback links an anomalous sea-surface temperature (SST) gradient to a sea level pressure (SLP) gradient that modulates trade-wind strength, creating a self-amplifying loop that drives event development. To explain the quasi-periodicity that Bjerknes feedback alone could not capture, Suarez and Schopf (1988), Battisti (1988), and Battisti and Hirst (1989) proposed the delayed oscillator model, in which equatorial Kelvin and Rossby waves and their boundary reflections carry the information needed to reverse the cycle. Jin (1997) subsequently framed ENSO through oceanic heat-content "recharging" and "discharging" in the equatorial warm pool, arguing that the build-up and release of warm water establish the necessary preconditioning for phase transitions.

### *b. Limitations of tropics-centric models and the search for external forcing*

Although these classical theories have substantially advanced our understanding, they primarily focus on coupled processes within the tropical Pacific and leave four persistent puzzles only partially resolved. **The event trigger problem:** classical models struggle to explain precisely why a given ENSO event begins or ends when it does. **Cycle irregularity:** ENSO's period varies between 2 and 7 years (McPhaden 1999), an irregularity that is difficult to reproduce with purely linear oscillator theories. **ENSO diversity:** observations reveal substantial diversity, including El Niño Modoki (Ashok et al. 2007) and the central-Pacific (CP) type (Kao and Yu 2009), whose formation mechanism differs from classical eastern-Pacific (EP) El Niño. **The spring predictability barrier:** ENSO forecast skill drops sharply during boreal spring, suggesting the influence of seasonally-dependent external factors that existing models underweight.

These limitations have prompted the community to look beyond the tropics. In the early twenty-first century, research began to explore connections between mid-to-high-latitude Pacific circulation anomalies and ENSO formation (Vimont et al. 2003; Anderson 2007; Yu and Kim 2011; Pegion and Selman 2017). Anderson (2007) found that SLP anomalies in the mid-latitude North Pacific, together with heat-content anomalies in the tropical western Pacific, are significantly related to El Niño triggering, although the proposed mechanism appears only weakly anchored to the annual cycle. These findings collectively suggest that a more complete ENSO theory must incorporate extratropical–tropical interactions into its core framework rather than treat them as peripheral modifiers.

### *c. A new hypothesis: the Pacific subtropical high as the primary forcing mechanism*

Building on an understanding of these limitations, Wang (2019) and Wang and Qin (2023) proposed a new theoretical framework whose core argument is that the fundamental driving force of the ENSO cycle originates in the extratropics, not the tropics. Specifically, the SLP anomalies in the Eastern Pacific Subtropical High (EPSH) belts of the Northern and Southern Hemispheres (NH and SH) constitute the higher-level forcing that modulates anomalous zonal winds in the equatorial Pacific; these wind anomalies, in turn, drive the development and evolution of equatorial SST anomalies. This perspective shifts the primary action center and forcing mechanism of ENSO from the interior of the tropical Pacific, as traditionally understood, to broader Pacific basin-scale interactions closely linked to the extratropics.

The theory constructs a clear causal chain: EPSH anomaly → equatorial Pacific zonal wind anomaly → equatorial SST anomaly. Within this framework, classical Bjerknes feedback is no longer the initiator of ENSO but a tropical response and amplification mechanism whose onset and intensity are regulated at a higher level by the extratropical EPSH. Yet the framework originally left three mechanistic claims essentially unvalidated: the proposed seasonal differentiation between a NH trigger and a SH sustainer had not been rigorously tested against the full observational record; the link between hemispheric synchrony and ENSO diversity was conjectural; and the Mirror-SIGMA (MSIGMA) termination mechanism, connecting mature ENSO anomalies back to the EPSH through the coupled Hadley–Walker circulation, had been described qualitatively but not quantified.

### *d. The solid-Earth dimension: LOD as an upstream geodetic precursor of ENSO initiation*

A parallel question concerns whether variability in Earth's rotation rate, quantified through Length of Day (LOD) observations, can influence ENSO initiation rather than merely responding to it. The ENSO-to-LOD response through atmospheric angular momentum (AAM) redistribution is well established. During ENSO events, reorganized wind and pressure fields redistribute AAM, which is exchanged with the solid Earth via friction and mountain torques, inducing subtle but measurable LOD anomalies (Rosen and Salstein 1983; Chao 1988; Dickey et al. 1992, 1994; Xu et al. 2023). The converse question, whether LOD variations can act as genuine precursors of ENSO, remains contested. Zheng et al. (2003) noted LOD anomalies preceding certain events. Weickmann et al. (1997) and Weickmann and Berry (2007) documented the sub-seasonal LOD/AAM variations organized by the Madden–Julian Oscillation (MJO; Madden and Julian 1971; Zhang 2005), whose westerly wind events can act as stochastic ENSO triggers (McPhaden 1999; Kessler 2001). However, these findings drew on few high-impact events, lacked a quantified atmospheric pathway, and could not separate the upstream precursor from the concurrent response (Zheng et al. 2003).

Here we bridge these two lines of inquiry. We provide the first systematic validation of the EPSH seasonal roles, the synchrony–diversity link, and the MSIGMA termination mechanism across 61 years and 30 diverse ENSO cycles, using rigorous lead–lag correlation and formal mediation analysis. We then extend the causal chain upstream, meaning earlier in time rather than upwind in the flow, to the solid Earth by extracting a stable sub-annual precursor signal from LOD variability that operates in phase with the annually modulated EPSH pacemaker. This combined analysis establishes that LOD not only responds to ENSO but also participates in modulating the probability of its initiation. Our findings reveal a genuinely bidirectional LOD–ENSO relationship and simultaneously construct a complete, statistically validated causal chain from solid Earth through extratropical atmosphere to tropical ocean. Table 1 summarizes the conceptual standing of this framework relative to classical theories.

**TABLE 1.** *Comparative analysis of ENSO theories. The EPSH-driven framework presented in this study is contrasted with classical ENSO theories across key dimensions.*

| Aspect | EPSH-Driven (This study) | Delayed Oscillator | Recharge–Discharge |
|---|---|---|---|
| **Primary forcing** | SLP anomalies in EPSH belts of both hemispheres, further modulated by LOD(0.5–1 yr) /seasons | Tropical internal wind-stress anomalies | Western Pacific warm water volume changes |
| **Forcing location** | Subtropical Pacific (20–40° N/S), central to eastern longitudes | Tropical Pacific | Tropical western Pacific |
| **Phase-locking** | Annual cycle modulates EPSH meridional migration, locking phase to year-end | Weak; mature phase timing often mismatches observations | Weak; recharge/discharge timescale not fully aligned with seasonal phase-locking |
| **ENSO diversity** | NH/SH EPSH synchrony determines EP vs. CP type | Difficult; focuses on a single oscillation mode | Difficult; focuses on basin-wide heat content changes |
| **Upstream solid-Earth link** | Quantified LOD(0.5–1 yr) → EPSH pathway; three-stage ridge engagement (Y−1 Aug, Y−1 Dec, Y0 May) fixing ENSO phase 14–17 months ahead | Not considered | Not considered |

### *e. Report objectives and structure*

This study pursues four linked goals. *First*, we validate the causal chain from the EPSH to the equatorial ocean through detailed statistical analysis across 61 years of observations. *Second*, we examine the asynchronous forcing of the NH and SH EPSH and the role of hemispheric synchrony in shaping ENSO diversity, using the contrasting 1991/92 and 1997/98 events as quantitative case studies. *Third*, we embed this core mechanism in the broader climate context by integrating solid-Earth rotation variability (LOD), intraseasonal tropical variability (MJO), and the annual cycle into a single hierarchical framework. *Fourth*, we extend the analysis to event termination by quantifying the role of the MSIGMA pattern and documenting its independent geodetic fingerprint in the LOD record.

The remainder of the paper is organized as follows. Section 2 describes the data, indices, pre-processing, and statistical methods. Section 3 presents the statistical evidence for the EPSH pacemaker and the central-Pacific action center. Section 4 examines the hemispheric roles, the contrasting Modoki and super El Niño case studies, and the El Niño–La Niña asymmetry. Section 5 reports the LOD spectral structure, the full-sample causal chain, and the mediation analysis that quantifies the upstream solid-Earth signal. Section 6 synthesizes the results into a unified hierarchical framework, section 7 discusses the implications for prediction, and section 8 concludes. Technical details are placed in appendix A and the supplemental material.

## 2. Data and methods

### *a. Data*

Monthly sea level pressure and surface zonal wind on a 2.5° latitude–longitude grid are from the NCEP/NCAR global reanalysis (Kalnay et al. 1996) for 1963–2023. Sea-surface temperature is from the NOAA Extended Reconstructed SST v5 on a 2° grid for the same period (Huang et al. 2017). Daily LOD values are from the IERS Earth Orientation Parameters C04 20 series (Bizouard et al. 2019), averaged to monthly means. The active analysis period for the LOD causal chain is January 1963 – December 2022 (720 months).

### *b. Index definitions and notation*

Twelve regional indices are defined as area-mean anomalies relative to the 1963–2023 monthly climatology; the suffix "a" denotes anomaly throughout. The EPSH belt in each hemisphere is first partitioned into three longitudinal sectors, western, central, and eastern (SLP_R1–R3 in the Northern Hemisphere, SLP_R4–R6 in the Southern; Table A1 of appendix A). Of these six candidates, the central sector in each hemisphere, SLP_R2 (20–30°N, 150–120°W) and SLP_R5 (20–30°S, 150–120°W), carries by far the strongest and most consistent relationship with the downstream wind and SST fields (Figs. 1 and 3); the other four sectors add no independent information once R2 and R5 are accounted for. These two indices are therefore used throughout as the working representation of the Northern and Southern Hemisphere EPSH, and for brevity we refer to them collectively as the EPSH boxes. The remaining indices used in the main analysis are their sum SLP_R2+R5, U_R1 (2.5°S–2.5°N, 160–180°E; central Pacific zonal wind), SST3 (5°S–5°N, 150–90°W; NINO3), and SST_R1 (2.9°S–2.9°N, 180–150°W; central Pacific SST). Full definitions of all 12 regional indices are listed in Table A1.

### *c. LOD pre-processing and spectral decomposition*

Monthly LOD anomalies are computed by subtracting the 1963–2022 monthly climatology. Six astronomical tidal periodicities (1.0, 0.5, 1/3, 0.25, 18.613, and 9.307 yr) are removed by least-squares fitting and subtraction, following Chao (1988) and Xu et al. (2023), yielding the de-tided LOD anomaly series (standard deviation 0.839 ms). This series is decomposed into five frequency bands using fourth-order zero-phase Butterworth bandpass filters (0–0.5 yr, 0.5–1 yr, 1–1.5 yr, 1.5–2 yr, 2–10 yr) with AR(15) autoregressive endpoint extension to suppress Gibbs-phenomenon ringing. Filter specifications and band variance contributions are summarized in supplemental Table S1 (the supplemental material), together with a seasonal bias test confirming the absence of spurious seasonality in the 0.5–1 yr band.

### *d. ENSO classification*

ENSO events are classified by the December–January–February (DJF) mean SST3a, computed for December of year Y, January of Y+1, and February of Y+1. The 15 highest DJF years are designated El Niño events (1965, 1972, 1976, 1982, 1986, 1987, 1991, 1994, 1997, 2002, 2003, 2006, 2009, 2015, 2018); the 15 lowest are La Niña events (1964, 1967, 1970, 1971, 1973, 1975, 1984, 1988, 1996, 1998, 1999, 2007, 2010, 2017, 2021).

### *e. Statistical methods*

All statistical significance tests reported are two-tailed. Significance thresholds for Pearson correlation coefficients are $|r| > 0.252$ (95%) and $|r| > 0.327$ (99%) for the EPSH analysis (N = 61 years, 1963–2023, d.f. = 59), and $|r| > 0.255$ (95%) and $|r| > 0.330$ (99%) for the LOD causal chain analysis (n = 59 years, 1964–2022, d.f. = 57; the years was extending to 2025 for the test of 2026 prediction). Composite analysis uses Welch's two-sample t-test applied to the Y−1 and Y0 portions of composites; the Y+1 portion is shown for qualitative reference only, because the LOD signal in Y+1 is partly driven by the ENSO event itself through the ENSO → AAM → LOD feedback. Mediation is assessed in the Baron–Kenny framework (Baron and Kenny 1986) implemented through partial correlations: the total effect, the paths into and out of each candidate mediator, and the direct residual path with the mediator held fixed are each tested directly, with no resampling required. A bootstrap product-of-coefficients analysis (5,000 resamples; Preacher and Hayes 2008) is retained as a corroborating check and reported in the supplemental material. For the LOD causal chain, the parametric thresholds above are supplemented by a phase-randomization Monte Carlo test: 5,000 surrogate LOD series are generated by randomizing the Fourier phases of the 0.5 to 1 yr band while preserving its full power spectrum, and hence its bandpass persistence; the A2 index is recomputed from each surrogate and empirical p-values are read from the resulting null distributions. This procedure requires no prior assumption about the effective degrees of freedom. A selection-aware variant charges the index for the choice of sampling window: each surrogate is allowed the same search over the nine one-month window variants (start February to April, end June to August) and the maximum |r| is retained, so that the observed statistic is compared against a null distribution that has undergone the identical search (the supplemental material). The implied effective sample size is approximately 50 (about 55 for the SLP links and about 42 for NINO3), with an empirical 95% threshold of approximately |r| = 0.27 to 0.30. A field-significance test applies the same surrogates to the full Pacific SLP correlation map: the observed fraction of the domain exceeding the local 5% threshold is 27%, against a surrogate 95th percentile of 17%, giving field significance P = 0.009.

## 3. The causal chain from extratropics to tropics: statistical evidence

This section systematically validates the causal chain from the extratropics to the tropics proposed by the EPSH-driven theory. We first demonstrate the key role of the central equatorial Pacific zonal wind (U_R1) as the direct oceanic driver, then establish the preceding role of the EPSH as a pacemaker, and finally show how the annual cycle acts as the metronome regulating the pacemaker's tempo. The ±12-month full lead–lag structure is documented first (Fig. 1) to motivate the 2-month representative lag used in subsequent monthly analyses.

### *a. Identifying the key forcing region: the primacy of central-Pacific zonal wind*

Traditional ENSO theories often treat tropical wind stress as a monolithic driver, whereas the EPSH-driven framework emphasizes the importance of wind fields in specific regions. To precisely locate the zonal wind area most influential in ENSO formation, we divide the equatorial Pacific into three regions: U_R1 near the dateline (160°E–180°), the central U_R2 (170°–150°W), and the eastern U_R3 (140°–110°W), all over 2.5°N–2.5°S. Computing the lagged correlations between zonal wind anomalies (Ua) in each region during spring (MAM), summer (JJA), and autumn (SON) with the subsequent winter (DJF) Niño3 SST anomaly (SST3DJFa) then reveals which region carries the strongest predictive value.

Table 2 clearly reveals the critical role of U_R1a. In the early development stage during spring, the correlation between U_R1a and subsequent winter SST3DJFa reaches 0.513, far higher than U_R2a (0.308) and U_R3a (0.053). U_R3a, the wind field closest to the Niño3 region, has essentially zero predictive signal in spring. As the seasons progress, correlations for U_R1a and U_R2a continue to strengthen, both exceeding 0.8 in autumn, indicating their important roles during ENSO development and maturity. The significant leading correlation of U_R1a in spring makes it the best single indicator for identifying early ENSO signals and external forcing.

**TABLE 2.** *Lagged correlations of seasonal zonal wind anomalies (Ua) in three equatorial Pacific regions with subsequent winter SST3DJFa, 1963–2023. Bold values denote significance at the 99% confidence level (|r| > 0.33). U_R1a shows the strongest spring predictive signal, identifying the central Pacific near the dateline as the primary action center.*

| Season | U_R1a vs. SST3DJFa | U_R2a vs. SST3DJFa | U_R3a vs. SST3DJFa |
|---|---|---|---|
| **Spring (MAM)** | **0.513** | 0.308 | 0.053 |

| Season | U_R1a vs. SST3DJFa | U_R2a vs. SST3DJFa | U_R3a vs. SST3DJFa |
|---|---|---|---|
| **Summer (JJA)** | **0.736** | **0.778** | 0.278 |
| **Autumn (SON)** | **0.862** | **0.898** | **0.467** |

This finding carries clear physical significance. The key action center driving ENSO SSTa is located in the central equatorial Pacific near the dateline. As research into ENSO diversity has deepened, the community has gradually recognized the importance of the central Pacific, particularly for CP-type ENSO or El Niño Modoki. Yet earlier work treated this as a property of one event type, whereas the present framework shows that the central Pacific is the initial and core response region to extratropical forcing in essentially all ENSO events, providing a new physical perspective for understanding the entire cycle.

***b. The higher-level driver: lagged impact of EPSH anomalies on equatorial winds***

Since U_R1a is the direct oceanic driver, the next question is: what controls the changes in U_R1a? The EPSH-driven framework points to the extratropical subtropical highs. To verify this higher-level driving relationship, we analyze the full lead–lag structure between SLP anomalies in the key EPSH regions and U_R1a (Fig. 1). Both SLP_R2a (NH EPSH central, 20–30°N, 150–120°W) and SLP_R5a (SH EPSH central, 20–30°S, 150–120°W) peak in negative correlation with U_R1a at +2 months ($r = -0.41$ and $-0.50$, respectively), while SLP_R5a peaks against SST3a at +3 months ($r = -0.46$), one month later than for U_R1a. This sequential ordering EPSH → U_R1a → SST3a establishes U_R1a as the mediating action center and motivates the 2-month representative lag used below.

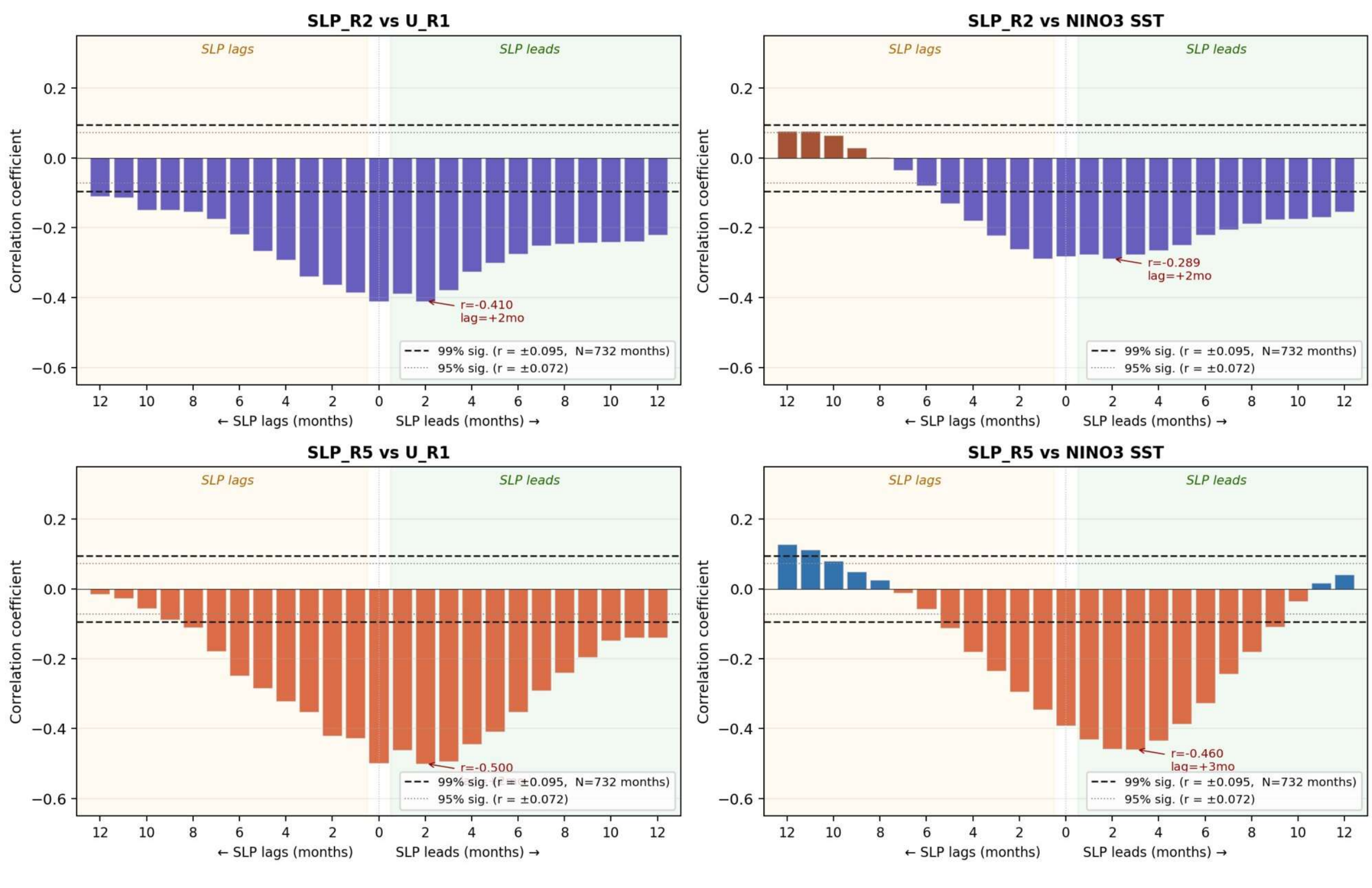

FIG. 1. Full ±12-month lead–lag correlation analysis between subtropical SLP and equatorial variables, 1963–2023 (N = 732 monthly anomaly values). Positive lags (right half, "SLP leads") represent the extratropical-to-tropical causal direction; negative lags (left half, "SLP lags") capture the reverse or feedback direction. (a) SLP_R2a versus U_R1a; the most negative correlation occurs at lag +2 months (r = −0.41), identifying a 2-month atmospheric adjustment timescale. (b) SLP_R2a versus SST3a; correlations are weaker and peak at a longer lead, consistent with the slower oceanic SST response. (c) SLP_R5a versus U_R1a; the SH EPSH shows a stronger and more persistent negative lead correlation than the NH (peak r = −0.50 at +2 months), confirming its role as the sustained amplifier. (d) SLP_R5a versus SST3a; the most negative correlation occurs at lag +3 months (r = −0.46), one month later than the peak for U_R1a, documenting the sequential ordering EPSH → U_R1a → SSTa. Dashed lines: 99% significance threshold; dotted lines: 95% threshold, both under the assumption of serial independence. Accounting for the lag-1 autocorrelations of the monthly series (Bretherton et al. 1999) gives an effective sample size of approximately 340 and adjusted 95% and 99% thresholds of |r| ≈ 0.11 and 0.14; the peak correlations of −0.41 to −0.50 exceed even the adjusted thresholds by a wide margin.

Using the 2-month lag motivated above, Fig. 2 shows the monthly evolution of lead correlations for SLP_R2a and SLP_R5a against three target variables. Three features emerge. **Significant negative correlation:** nearly all values in Fig. 2a surpass the 99% significance level (|r| > 0.33, d.f. = 59), confirming a robust inverse relationship; a weaker EPSH (negative SLPa) is

associated with subsequent westerly wind anomalies (positive Ua) in the central equatorial Pacific. **Asynchronous forcing:** the correlation curves are strikingly out-of-phase. SLP_R2a peaks in boreal winter–spring with its deepest trough in February ($r = -0.59$), when the NH high reaches maximum equatorial proximity. SLP_R5a peaks in boreal summer–autumn, reaching $r = -0.77$ in September, the strongest single signal in the dataset. **Seasonal dominance:** this out-of-phase relationship highlights the complementary and seasonal nature of the hemispheric influences. The Northern Hemisphere acts primarily as a trigger in winter–spring, while the more powerful and persistent Southern Hemisphere acts as a sustainer and amplifier in summer–autumn. Lagged correlation curves for SLP and equatorial SST (Fig. 2b) are broadly similar but slightly weaker, confirming that the wind anomaly is the more direct response. Correlations involving the easternmost U_R3a (Fig. 2c) are generally less significant, reinforcing the primacy of the central Pacific action center.

**FIG. 2.** Monthly evolution of 2-month lead correlations between subtropical SLP and equatorial variables. Three panels show Pearson correlation coefficients (y-axis) for each calendar month (x-axis), where SLP in a given month is correlated with the target variable two months later. (a) SLP_R2a (NH EPSH; blue) and SLP_R5a (SH EPSH; red) versus U_R1a. SLP_R2a reaches its deepest negative correlation in February ($r = -0.59$), identifying the NH EPSH trigger phase. SLP_R5a reaches its peak in September ($r = -0.77$), identifying the SH EPSH sustainer phase. (b) Same SLP indices versus SST_R1a (central Pacific SST); broadly similar to (a) but slightly weaker. (c) SLP_R2a and SLP_R6a (SH EPSH eastern sector) versus U_R3a (eastern equatorial wind); correlations are generally weaker, confirming U_R1 near the dateline as the primary action center. Horizontal dashed and dotted lines indicate the 99% ($|r| = 0.327$) and 95% ($|r| = 0.252$) significance

thresholds, respectively (N = 61 years, d.f. = 59).

### *c. The role of the annual cycle: the metronome modulating the pacemaker*

If the EPSH is the pacemaker of ENSO, the annual cycle is the metronome that sets its tempo. ENSO events exhibit a strong seasonal phase-locking feature, with most events peaking near the end of the calendar year (Rasmusson and Carpenter 1982; Stein et al. 2014). Traditional theories have struggled to provide a convincing physical explanation for this, whereas the EPSH-driven theory offers a direct answer. Seasonal changes drive the regular meridional migration of the global atmospheric circulation, including the subtropical high-pressure belts; according to Wang (2019), the subtropical high belts as a whole move northward by about 5° of latitude in boreal summer compared with boreal winter. This view of the annual cycle as a key modulator aligns with perspectives that treat ENSO and the annual cycle as an interacting combined mode (Stuecker et al. 2013).

Although the annual cycle is not the direct pacemaker of the ENSO system, its key role is seasonal modulation of the strength and equatorial influence of the NH and SH EPSH. In late winter and early spring, although the NH EPSH anomaly is comparatively weaker and shorter-lived than the SH EPSH, it often plays the crucial role of trigger, capable of initiating an ENSO event. As the cycle progresses into the following summer, the EPSHs migrate northward; the SH EPSH, now positioned closer to the equator, becomes the dominant driver of the equatorial trade winds, fostering conditions conducive to either La Niña or sustained El Niño development depending on the sign of the anomaly. Because the SH EPSH generally possesses greater size and volume than its NH counterpart, its prolonged impact on the equatorial trade winds during summer and autumn is a critical factor contributing to the peak of ENSO events toward year-end.

Figure 3 provides the spatial complement to this seasonal narrative. It shows the correlation between SLP in March and U_R1a in subsequent months of the same year. A prominent feature is the persistent, symmetric, large-scale negative correlation area on both the northern and southern sides of the eastern Pacific mid-to-high latitudes. The pattern strengthens from May to September, with the most coherent and spatially extensive correlations in the SH subtropical region during September, consistent with the SH EPSH sustainer role reaching peak equatorial influence. The symmetric pattern in March highlights a critical window where the system is particularly sensitive to coordinated forcing from both hemispheres: if the NH and SH EPSH belts weaken coherently in early spring, they provide an exceptionally strong and lasting initial push for ENSO onset. This spatial evidence independently corroborates the monthly curves in Fig. 2 and anchors the case-study narratives presented in the next section.

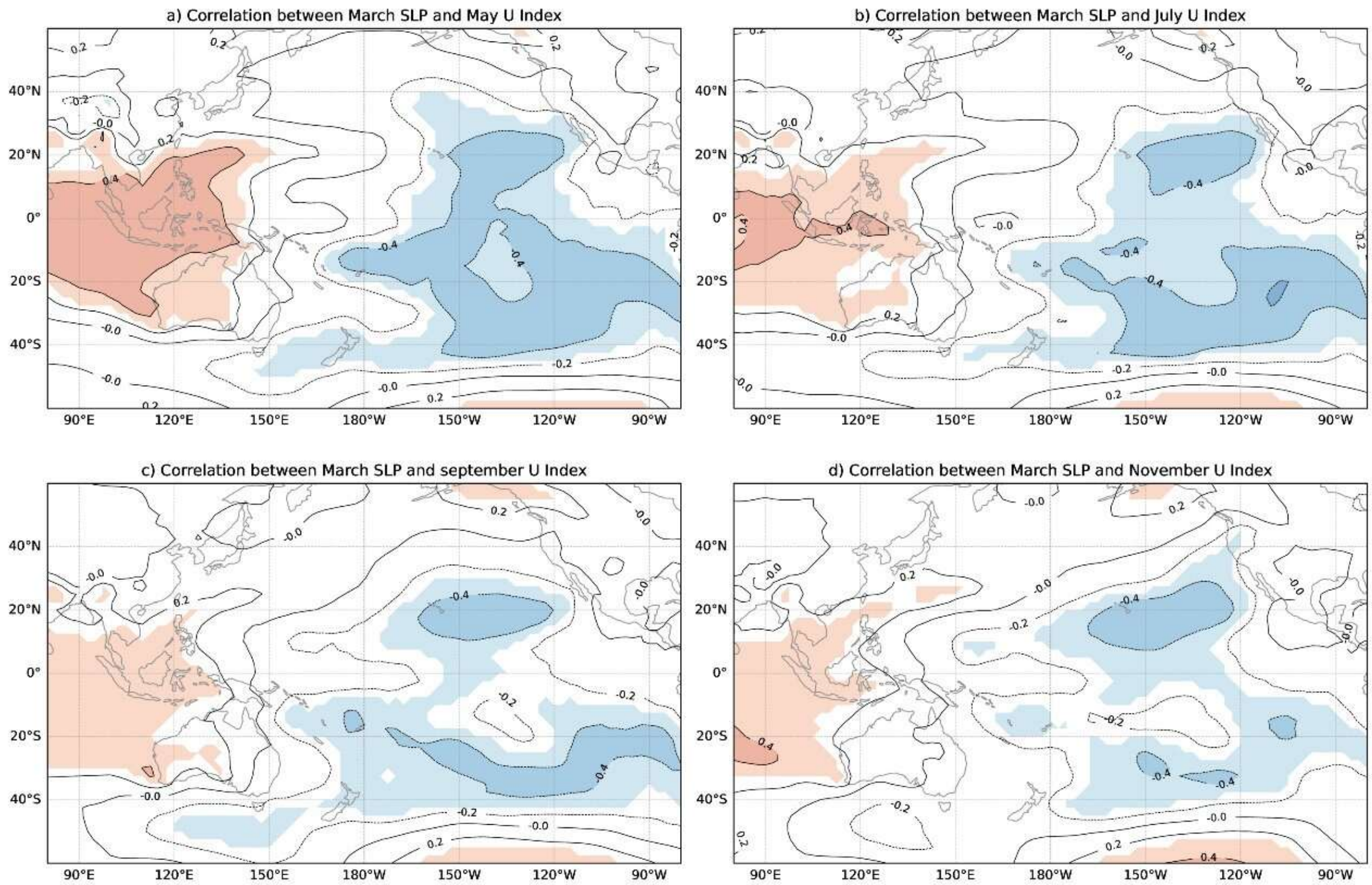

**FIG. 3.** Spatial correlation maps between SLPa in March and U_R1a in subsequent months, 1963–2023. (a) May, (b) July, (c) September, (d) November. Shaded areas are significant at the 95% level. Contour interval: 0.1; dashed contours indicate negative values. A persistent, symmetric, large-scale negative correlation pattern is visible in both hemispheric subtropical regions (20–40°N and 20–40°S) of the eastern Pacific, centered on 150–120°W. This pattern strengthens from May to September, with the most coherent and spatially extensive correlations in the SLP_R5 region during September (panel c), consistent with the SH EPSH sustainer role reaching peak equatorial influence. The symmetric March pattern highlights a critical window where coordinated weakening of both hemispheric EPSH belts provides exceptionally strong initial forcing for ENSO onset.

## 4. A tale of two hemispheres: asynchronous forcing, seasonal roles, and ENSO diversity

The NH and SH EPSH systems do not always vary in sync. Their complex interactions, particularly their different roles in different seasons, are key to understanding the diversity of ENSO events (EP vs. CP El Niño) and the asymmetry between El Niño and La Niña. Importantly, the simultaneous correlation between SLP_R2a and SLP_R5a is statistically insignificant during JJA (Table 3), corroborating the unique role played by the stable SH EPSH in shaping ENSO during summer and providing the physical prerequisite for hemispheric asynchrony to express itself as event diversity.

**TABLE 3.** *Monthly Pearson correlation between SLP_R2a and SLP_R5a, 1963–2023 (n = 61 per month). Bold values denote significance at the 99% confidence level ($|r| > 0.33$); italic values denote 95% ($|r| > 0.25$); other values are not significant. The JJA correlations are not significant, confirming independent hemispheric operation during summer.*

| Month | Jan | Feb | Mar | Apr | May | Jun | Jul | Aug | Sep | Oct | Nov | Dec |
|---|---|---|---|---|---|---|---|---|---|---|---|---|
| **r(R2a,R5a)** | **0.364** | **0.467** | 0.220 | *0.259* | *0.277* | 0.174 | *0.272* | 0.105 | **0.425** | **0.528** | *0.256* | *0.319* |

### *a. The winter–spring trigger and the summer–autumn sustainer*

As established in Fig. 2a, the NH EPSH plays the role of a trigger or an early warning signal. Its influence, though significant, is shorter-lived than its SH counterpart. This explains why earlier studies (Anderson 2007) identified the winter SLPa in the North Pacific as a predictor for ENSO, but the signal alone could not guarantee that a full El Niño event would necessarily occur. In stark contrast to the trigger role of the NH EPSH is the sustaining and amplifying role of the SH EPSH. The physical basis for this is that the SH subtropical high is climatologically stronger than its NH counterpart; during JJA, its position is closest to the equator, maximizing its potential influence. The SH EPSH is therefore not only the sustainer of the ENSO cycle but also its amplifier. An initial NH disturbance, if met and amplified by strong and persistent SH forcing during summer and autumn, can develop into a significant ENSO event affecting the entire Pacific basin. Conversely, if the SH forcing fails to take over, the initial NH signal will gradually dissipate and the event will be stillborn. This two-stage forcing mechanism of NH trigger followed by SH sustainer is a core component of the new theory and is essential for explaining ENSO diversity. One qualification is worth stating plainly, because it bears on how the sequence begins rather than how it culminates: in the LOD-driven precursor chain (section 5b), it is the Southern Hemisphere ridge that responds first, within a month of the A2 window closing, while the Northern Hemisphere ridge is still inactive; only by the following spring does the Northern Hemisphere ridge assume the dominant role described above. The NH-trigger, SH-sustainer division therefore describes the concurrent, mature-season roles the two ridges play once the annual cycle has carried the highs into their respective active windows, not the order in which the precursor first engages them; the SH ridge's early involvement is a supporting contribution that should not be read out of the picture.

### ***b. A case study in ENSO diversity: the asynchronous 1991/92 vs. the synchronous 1997/98 event***

#### *1) The 1991/92 event: a product of asynchronous and intermittent forcing*

The 1991/92 El Niño is widely classified as close to El Niño Modoki (Timmermann et al. 2018). Figure 4a illustrates the seasonal evolution of equatorial SSTa, surface zonal wind anomalies, and extratropical SLP anomalies from DJF 1990/91 through DJF 1991/92, making the hemispheric forcing signals directly readable from the maps; the corresponding monthly time series of the same five variables are shown in Fig. 5a.

Three decisive stages can be identified. **Initial phase (late 1990–early 1991):** both the NH EPSH (SLP_R2 region, 20–30°N, 150–120°W) and the SH EPSH (SLP_R5 region, 20–30°S, 150–120°W) showed initial negative SLP anomalies (Fig. 4a, DJF 1990/91 panel; Fig. 5a), driving westerly Ua and a preliminary rise in central Pacific SSTa; the system was primed for an event. **Critical turning point (spring 1991):** SLP_R2a shifted sharply positive while SLP_R5a remained negative (Fig. 4a, MAM 1991 panel; Fig. 5a, spring 1991). This opposing signal actively suppressed development: U_R1a stagnated, eastward warm-water advection from the Western Pacific Warm Pool was impeded, and warming was confined to the central Pacific while growth of SST3a remained limited; the hallmark of a Modoki event. **Re-intensification (summer/autumn 1991):** SLP_R2a turned negative again (Fig. 4a, JJA and SON 1991 panels; Fig. 5a), becoming synchronous with the persistently weak SLP_R5a, but the critical spring window for eastward propagation had already passed.

Tropics (|lat|≤15°, shaded band): SSTa shading + Ua contours (green, 0.8 m $s^{-1}$, dashed easterly) • Extratropics (|lat|>15°): SLPa contours (black, 0.4 hPa, dashed negative) • green boxes = R2/R5

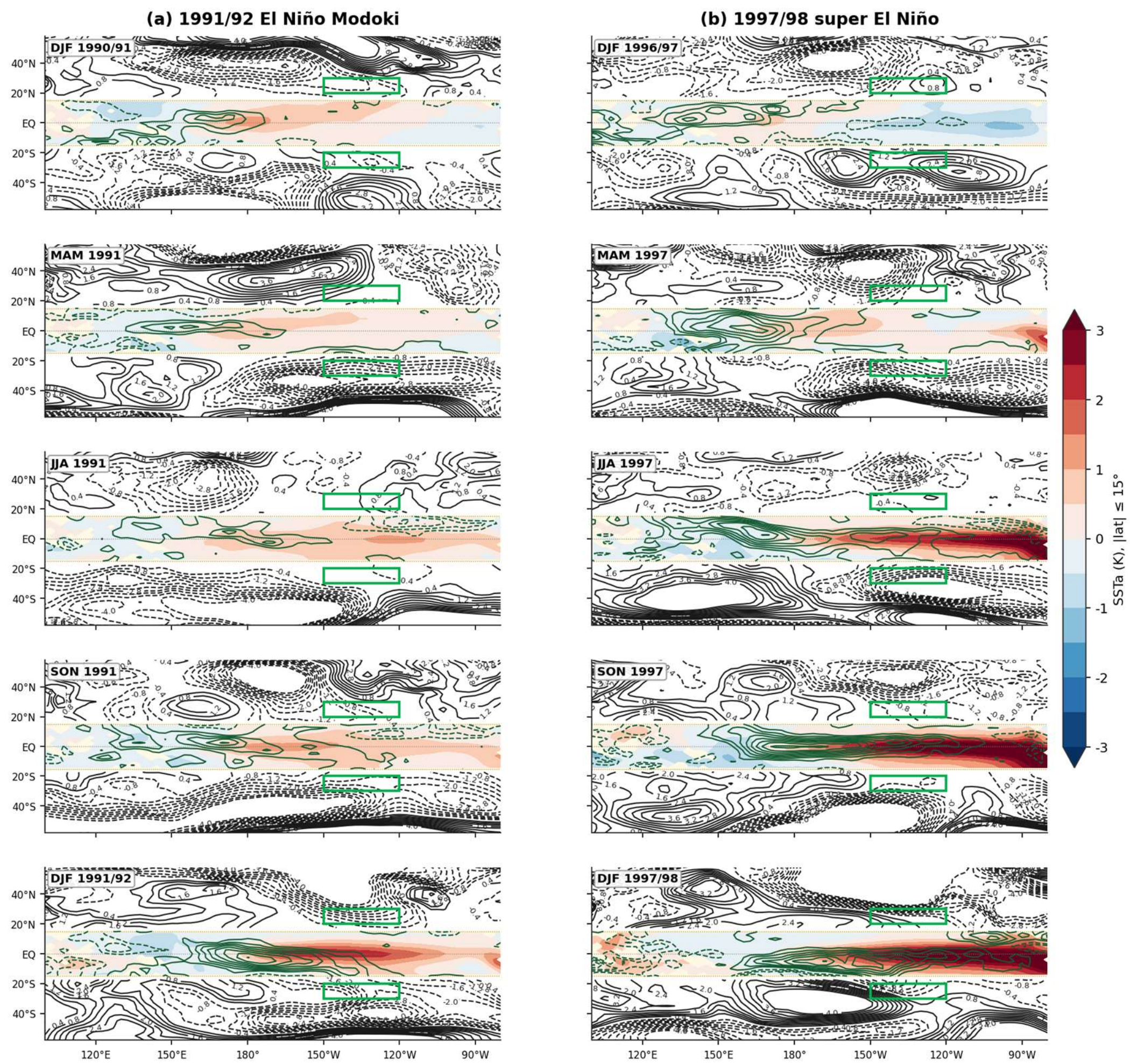

FIG. 4. Seasonal spatial distributions of anomalies for two contrasting ENSO case studies. Each case shows five seasonal panels covering DJF through the following DJF. Equatorial SSTa is shown as color shading (K; 95% significance), surface Ua as contour lines over the equatorial band (green, interval 0.8 m $s^{-1}$; dashed: easterly), and extratropical SLPa as contour lines over 15°–60° in both hemispheres (interval 0.4 hPa; dashed: negative). (a) The 1991/92 El Niño Modoki (five panels: DJF 1990/91, MAM 1991, JJA 1991, SON 1991, DJF 1991/92). The critical feature is the NH EPSH positive SLPa reversal visible in MAM 1991, co-occurring with persistent negative SLP_R5a in the SH. This asynchronous forcing suppresses equatorial westerly anomalies and confines warm SSTa to the central Pacific. (b) The 1997/98 super El Niño (DJF 1996/97, MAM 1997, JJA 1997, SON 1997, DJF 1997/98). Simultaneous negative SLPa in both hemispheric EPSH regions persists throughout the year with no spring reversal. Progressive basin-wide warming propagates eastward across successive seasons, culminating in warming exceeding +3°C across the equatorial Pacific by SON 1997 and DJF 1997/98.

*2) The 1997/98 super El Niño: a textbook case of synchronous and sustained forcing*

In stark contrast is the 1997/98 super El Niño (Fig. 4b), whose monthly five-variable evolution is shown in Fig. 5b. Unified onset (late 1996–spring 1997): from late 1996, EPSH anomalies in both hemispheres began a profound, persistent, and fully synchronous decline (Fig. 4b, DJF 1996/97 and MAM 1997 panels; Fig. 5b, late-1996 values). The NH trigger was itself strong and acted in concert with the SH signal, creating ideal preconditioning for explosive onset. Amplification phase (summer/autumn 1997): SLP_R5a became exceptionally strong and persistent (Fig. 4b, JJA and SON 1997 panels; Fig. 5b), taking over as sustainer and amplifier. Deeper mechanism: the synchronicity created a "perfect storm" that allowed Bjerknes feedback to activate at maximum intensity, sweeping warm water across the equatorial Pacific and producing basin-wide warming. U_R1a reached approximately +7 m $s^{-1}$ in October 1997 and SST3a peaked near +3.4°C shortly thereafter (Fig. 5b).

These two cases demonstrate that the difference between a Modoki and a canonical super El Niño does not stem from distinct physical mechanisms but from different operating modes of the same mechanism. The contrast between Figs. 5a and 5b is visually striking: the 1991/92 event shows the sign-reversal of SLP_R2a in spring 1991 and the stagnation of U_R1a, whereas the 1997/98 event shows coordinated, sustained negative SLPa in both hemispheres driving a persistent westerly U_R1a. The mode is determined by the synchronicity, persistence, and intensity of NH/SH EPSH forcing; when asynchronous, development tends toward a Modoki, when synchronous and sustained, the tropical feedback is fully unleashed. Hemispheric synchrony is in principle observable from EPSH regional indices in near-real time.

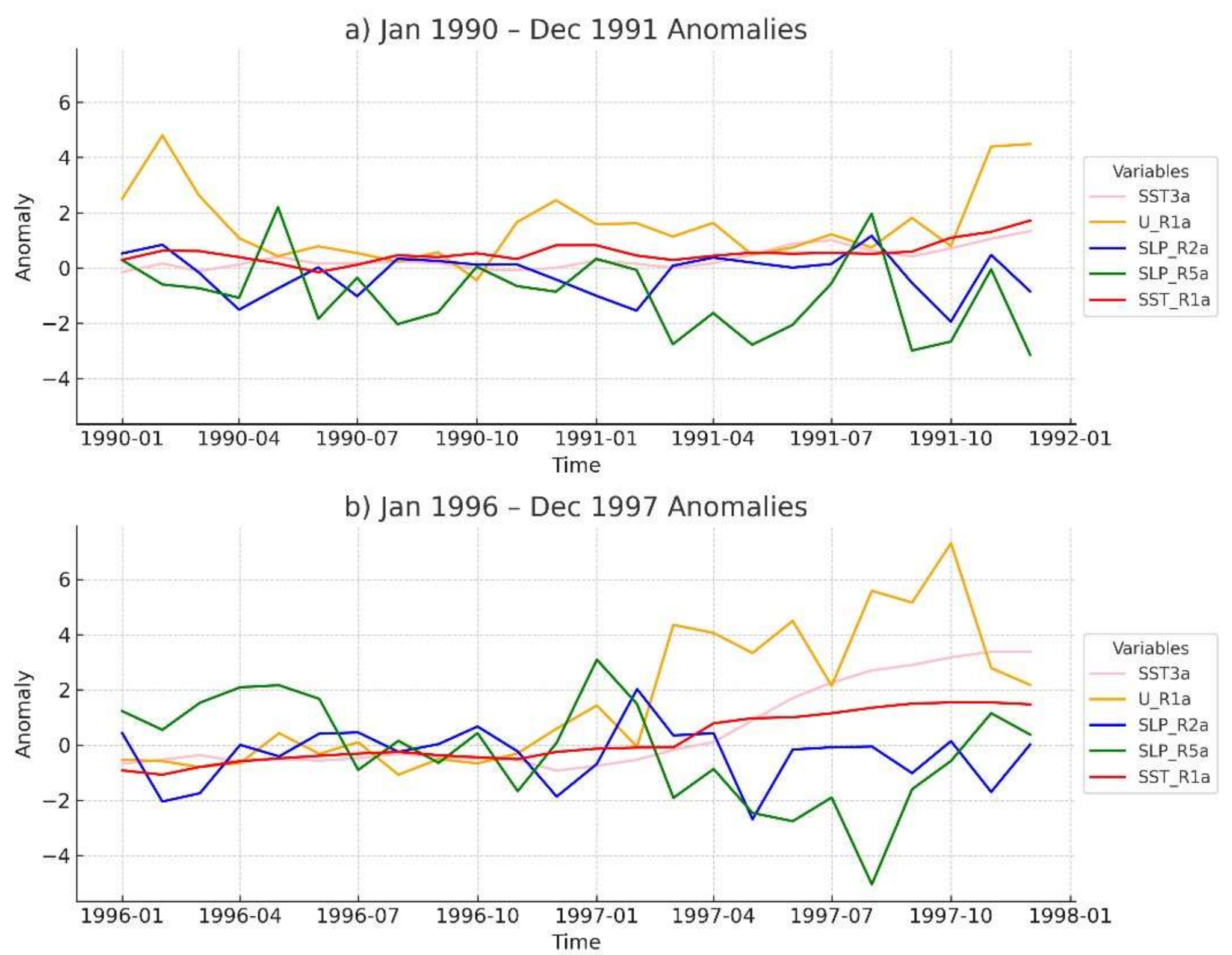

**FIG. 5.** Monthly evolution of key variables for the two contrasting ENSO case studies. (a) The 1991/92 El Niño Modoki, January 1990 to December 1991. (b) The 1997/98 super El Niño, January 1996 to December 1997. Both panels show the same five anomaly series: SST3a (light red), U_R1a (orange), SLP_R2a (blue), SLP_R5a (green), and SST_R1a (dark red). Units: SSTa in °C, U_R1a in m $s^{-1}$, SLPa in hPa. In (a), note the spring 1991 reversal of SLP_R2a from negative to positive while SLP_R5a remains negative; this asynchronous forcing is accompanied by weak, stagnating U_R1a and only modest SST3a development, confining warming to the central Pacific (Modoki signature). In (b), both SLP_R2a and SLP_R5a remain negative and broadly synchronous from late 1996 through autumn 1997, U_R1a develops a sustained strong westerly anomaly peaking near +7 m $s^{-1}$ in October 1997, and SST3a rises monotonically to approximately +3.4°C by the end of 1997 (super El Niño signature). The side-by-side comparison makes visually explicit how hemispheric synchrony discriminates event type.

### *c. Implications for ENSO asymmetry: comparing El Niño and La Niña development*

The evolution of El Niño and La Niña is not a simple linear mirror reversal. Besides the asymmetry in intensity, there are also subtle differences in their seasonal phase-locking characteristics. Composite evolution of the key variables across the 15 El Niño and 15 La Niña events (section 5b; see the SLP_R5a curves in Fig.7c) reveals three systematic asymmetries. Forcing timing: during El Niño development, the most significant weakening of the SH EPSH (negative SLP_R5a) occurs in September; during La Niña, the most significant strengthening of SLP_R5a peaks one month earlier, in August (Fig.7c). Response amplitude: anomalies are generally larger during El Niño events; the peak U_R1a and SST_R1a are stronger in the El Niño composite than their easterly/cold counterparts in the La Niña composite. Hemispheric contribution: in both event types, the SH forcing (SLP_R5a) shows a larger and more sustained anomaly during the crucial development phase than the NH forcing (SLP_R2a).

The one-month difference in the SH peak forcing, though seemingly small, reveals a key difference in the seasonal rhythm between the two event types. During La Niña, the peak strengthening of SLP_R5 occurs nearly simultaneously with the SH EPSH reaching its northernmost climatic position, consistent with direct annual-cycle control. During El Niño, the peak weakening of SLP_R5 is delayed by one month relative to the SH EPSH climatic position. The most plausible interpretation is that Bjerknes feedback during El Niño onset imposes a southward forcing on the SH EPSH through air–sea interaction: weaker trades diminish the barrier to eastward warm-water advection, and the cumulative effect produces the observed one-month lag. These findings imply that La Niña development tends to follow a more linear, dynamically straightforward pathway than El Niño, and that Bjerknes feedback may play a smaller role in La Niña formation.

## 5. Upstream from the subtropical highs: the LOD sub-annual precursor

Having established the EPSH pacemaker and its seasonal orchestration by the annual cycle, we now turn upstream to ask what modulates EPSH anomalies themselves. Large-scale variations in atmospheric angular momentum (AAM) and Length of Day (LOD) provide a planetary-scale diagnostic tool for ENSO and offer insights through the lens of global momentum conservation. The well-established ENSO-to-LOD pathway operates through AAM redistribution during mature events (Rosen and Salstein 1983; Chao 1988). Less well characterized is the reverse direction: whether LOD variability in specific frequency bands can serve as a genuine upstream precursor. We address this question through spectral decomposition, lead–lag correlation, and formal mediation analysis.

### *a. LOD spectral structure: separating the upstream precursor from the downstream response*

The de-tided LOD anomaly series is decomposed into five frequency bands using fourth-order zero-phase Butterworth bandpass filters (see section 2 and appendix A for filter specifications).

The resulting bands and their variance contributions are shown in Fig.S1. The 0–0.5 yr component (1.1% of variance) reflects intra-seasonal MJO-related AAM fluctuations; the 0.5–1 yr band (0.6%, dominant period about eight months) and the 2–10 yr band (3.2%) are the primary focus for connections to ENSO.

Twenty-four-month composites of all five bands for El Niño and La Niña events (Fig.6) reveal that four of five bands show near-mirror-image cold/warm polarity approaching 180° out-of-phase. The composite waveforms widen progressively from the 0.5–1 yr band (one visible cycle) to the 2–10 yr band (which converges to the SST3 evolution). This spectral hierarchy is diagnostically crucial: the 2–10 yr band reflects the well-established ENSO → LOD response through AAM redistribution, whereas the 0.5–1 yr band divergence is already visible in Y−1, before ENSO develops, identifying it as an upstream precursor whose positive phase peak implies conditions favorable for EPSH weakening and oceanic Kelvin wave generation.

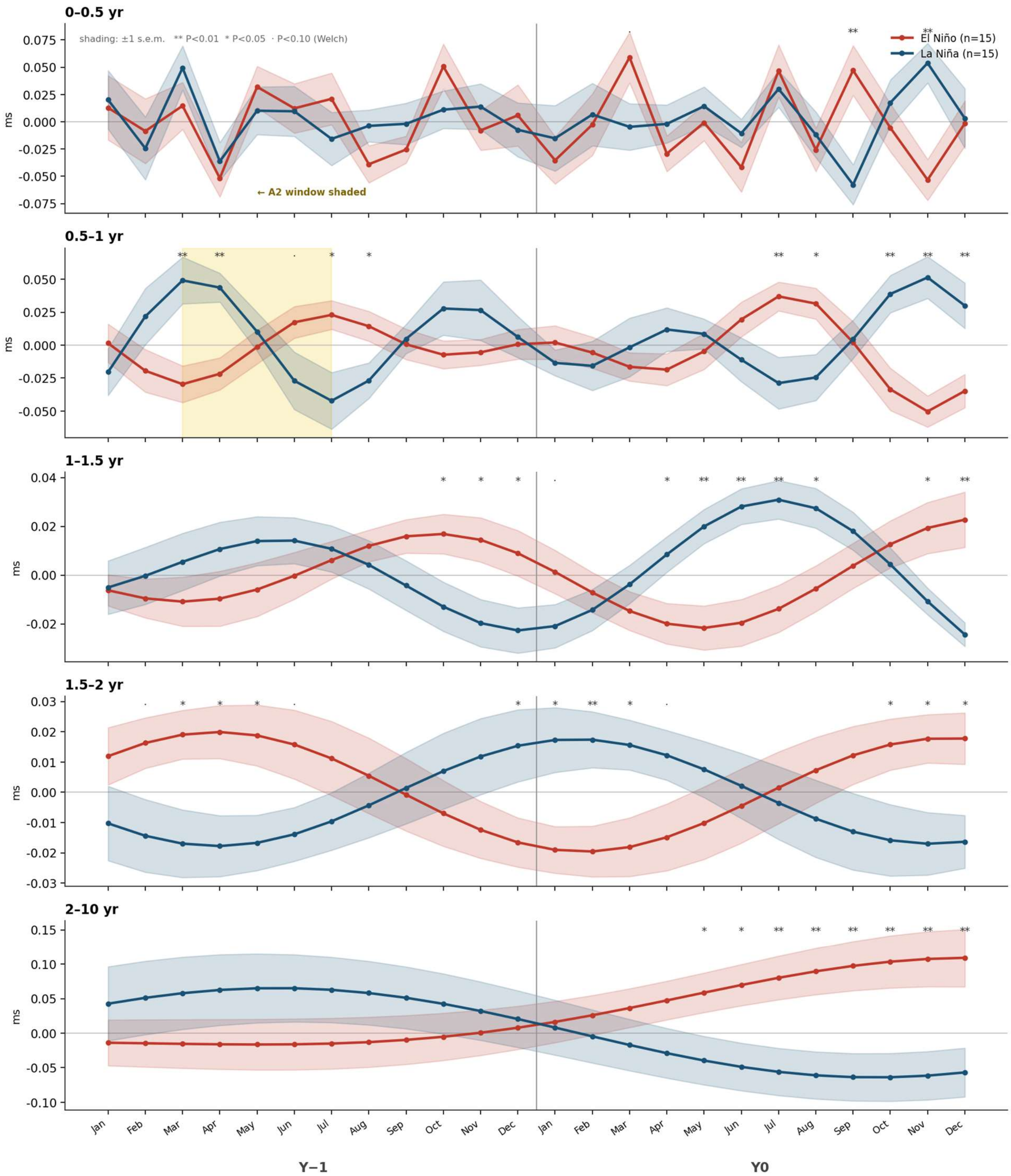

FIG. 6. Twenty-four-month LOD spectral composites for El Niño and La Niña events. Each panel shows the composite mean of one bandpass-filtered LOD component for El Niño (red, n = 15) and La Niña (blue, n = 15), spanning Y−1 January through Y0 December. A vertical line separates the precursor year (Y−1) from the event year (Y0). Shading around composite means: ±1 s.e.m. Significance symbols: ** $P < 0.01$, * $P < 0.05$, · $P < 0.10$ (Welch t-test, Y−1 and Y0 only). (a) 0–0.5 yr high-pass component: captures intra-seasonal MJO-related AAM fluctuations; no systematic cold/warm polarity. (b) 0.5–1 yr sub-annual component: the upstream precursor signal

identified in this study. Near-mirror-image polarity is already visible in Y−1, before ENSO develops. (c) 1–1.5 yr transition band and (d) 1.5–2 yr quasi-biennial band: both show a smooth, near-mirror-image oscillation between El Niño and La Niña composites that is already fully developed at the start of Y−1, rather than emerging within the precursor year as the 0.5–1 yr band does. A spring-to-summer index defined on each band in the same way as A2 correlates strongly with the previous winter's NINO3 ($r = +0.42$ for 1–1.5 yr, +0.39 for 1.5–2 yr, both $P < 0.01$) but not with the coming winter's ($|r| < 0.13$, not significant); read together with (e), these two bands are more plausibly the longer-period tail of the classical ENSO → LOD response, carrying the phase legacy of the preceding event, than an independent forward precursor for the event ahead. (e) 2–10 yr ENSO-band interannual component: the composite waveform has broadened to trace a slowly evolving signal that closely parallels SST3 evolution, reflecting the ENSO → LOD response. The spectral hierarchy across (b) to (e), progressive widening from one visible cycle to a broad ENSO-timescale signature, distinguishes the upstream precursor (0.5–1 yr band) from the downstream ENSO response (2–10 yr band).

### *b. The phase-alignment mechanism: two rhythms, one signal*

The stability of the 0.5–1 yr LOD precursor across 30 diverse ENSO events requires physical explanation. The EPSH system operates on an annual rhythm while the LOD (0.5–1 yr) oscillation has a dominant period of about eight months (the supplemental material; for the spectral character of the sub-annual AAM band see Weickmann et al. 2000); the two cycles are not commensurate, and their relative phase varies from year to year. ENSO initiation is favored when the positive LOD phase (Earth decelerating, westerly AAM tendency) coincides with the annual window of maximum EPSH equatorial influence. The 30-event composite captures precisely this phase alignment, which is why monthly-resolution signatures emerge systematically despite the sub-annual LOD periodicity.

This phase-alignment view leads to a natural index: the LOD amplitude A2 = LOD(Y−1, March) − LOD(Y−1, July) measures the spring-to-summer half-cycle amplitude of the LOD (0.5–1 yr) oscillation in the precursor year. A large negative A2 indicates that LOD rose substantially from March to July, reflecting a strong positive LOD phase during boreal summer when Earth deceleration transfers angular momentum to the atmosphere, enhancing westerly tendency and predisposing the EPSH toward weakening. The stability of this signal across 30 diverse ENSO events and its phase-locked operation with the annual cycle distinguish it from stochastic noise and establish it as a genuine sub-annual precursor embedded within the annually modulated pacemaker system. A five-variable composite extending the LOD analysis to include SLP_R2a, SLP_R5a, and U_R1a (Fig.7) confirms the sequential timing of the complete causal chain: LOD divergence precedes SLP divergence by several months, and U_R1a responds approximately 1–2 months after the SLP signal.

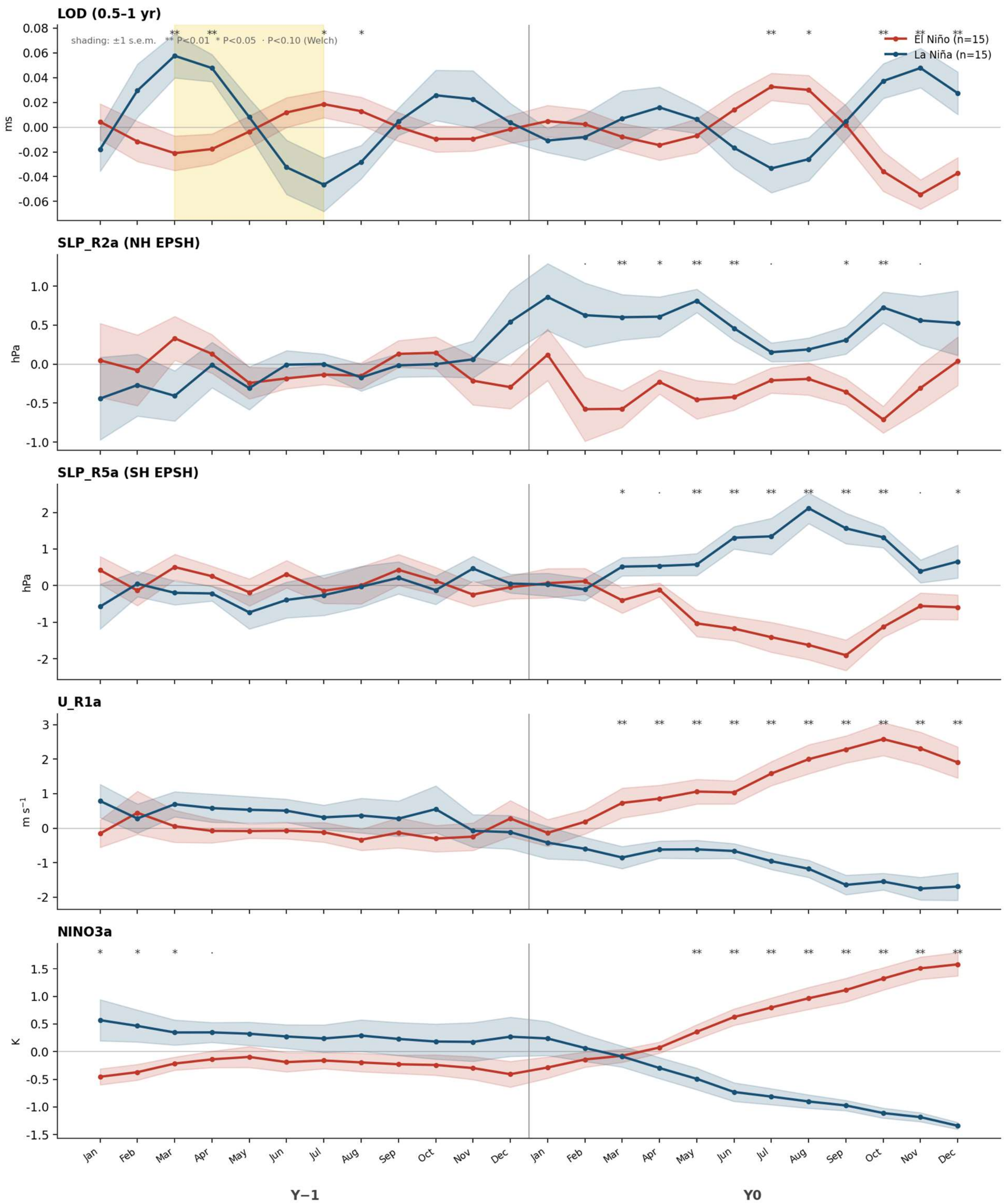

FIG. 7. Five-variable 24-month composite (Y−1 January to Y0 December). n = 15 + 15 events. El Niño: red. La Niña: blue. Shading: ±1 s.e.m. Significance: ** P < 0.01, * P < 0.05, · P < 0.10 (Welch t-test, Y−1 and Y0 only). (a) LOD (0.5–1 yr band; ms). El Niño and La Niña composites

diverge significantly from Y−1 March, with El Niño years more negative in boreal spring (Y−1 March–April) and more positive in summer (Y−1 June–August); this rising March-to-July swing is precisely the half-cycle measured by A2, consistent with an upstream precursor role. (b) SLP_R2a (hPa); significant divergence emerges from Y0 January–February. (c) SLP_R5a (hPa); similar pattern but with stronger and more sustained divergence through Y0 summer–autumn, consistent with the SH sustainer role. (d) SLP_R2+R5a (combined subtropical ridges, left axis) overlaid with U_R1a (right axis, dashed). The combined SLP index shows the strongest divergence. U_R1a responds approximately 1–2 months later, confirming the sequential EPSH → U_R1a ordering. (e) SSTa (°C); significant divergence from approximately Y0 March and peak separation in Y0 October. This panel documents the final step from atmospheric forcing to oceanic SST response.

### *c. Full-sample causal chain and mediation analysis*

To test the LOD–EPSH link across the full observational record, we compute lead–lag correlations of A2 against atmospheric and oceanic targets over 1964–2022 ($n = 59$ years; effective sample size $\approx 50$ by phase-randomized Monte Carlo). The subtropical highs are engaged at three successive stages, detailed in Fig. 8: the Southern Hemisphere ridge responds within a month of the window closing (Y−1 August), the eastern-Pacific system becomes coherent across the equator by Y−1 December, and the Northern Hemisphere ridge dominates a poleward-displaced structure by Y0 May. The May EPSH anomaly in turn drives U_R1a in Y0 June ($r = -0.47$, $P = 0.0002$) and SST3a in Y0 SON ($r = -0.53$, $P < 0.0001$), a sequential timing consistent with a forward-propagating causal chain rather than a simultaneous common-forcing response. Scatter plots disaggregated by ENSO category (Fig. S2) confirm the chain correlations are not driven by a few extreme events (supplemental Table S3), and the event-by-event relationship, together with the role of phase drift, is displayed in Fig. 9 and taken up in the discussion.

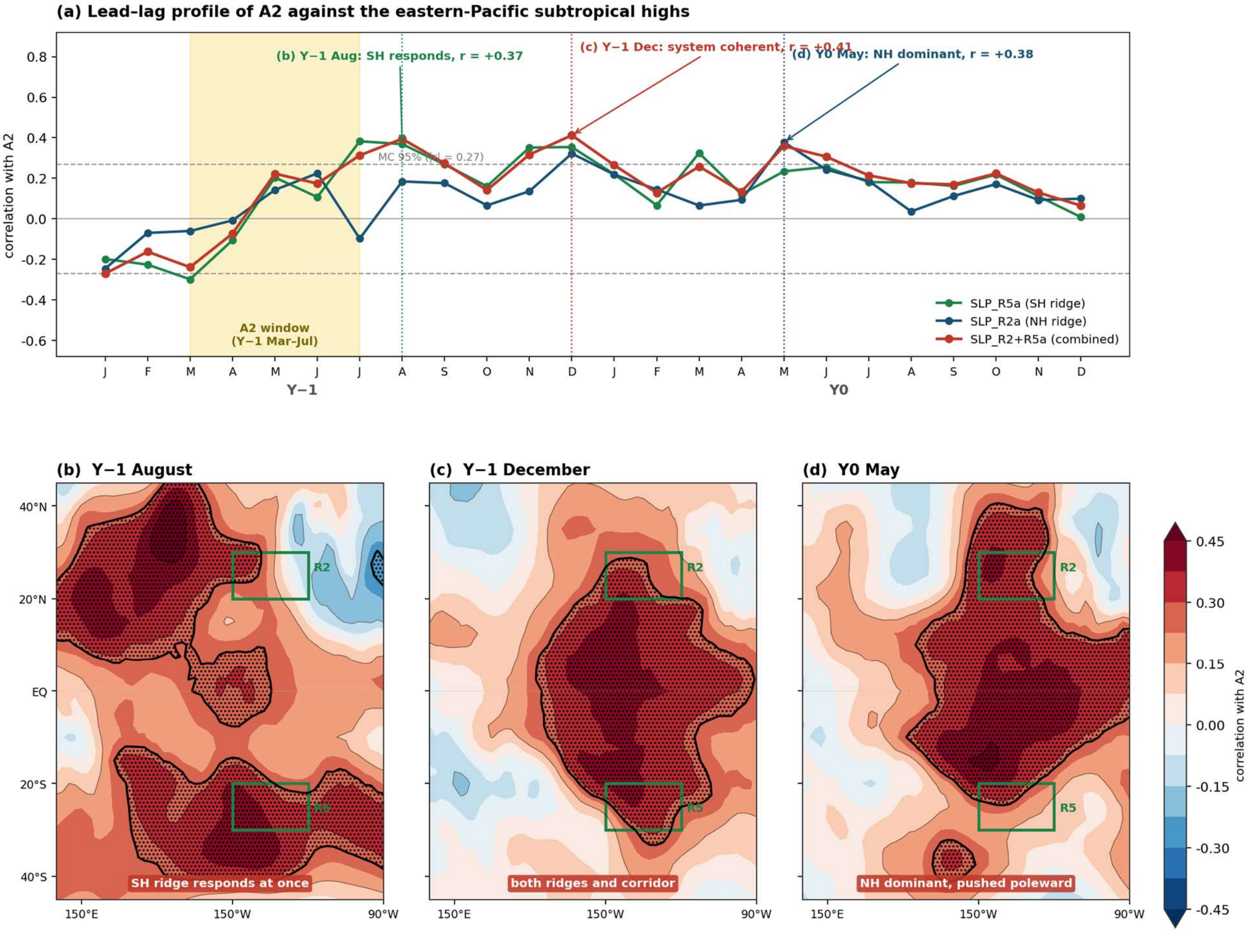

FIG. 8. When and where the sub-annual angular-momentum signal reaches the eastern-Pacific subtropical highs. A2 = LOD(Y−1 March) − LOD(Y−1 July), the spring-to-summer half-cycle amplitude of the LOD(0.5–1 yr) oscillation in the precursor year; n = 59 years (1964–2022). (a) Lead–lag correlation of A2 with SLP_R2a (blue, Northern Hemisphere ridge anomaly), SLP_R5a (green, Southern Hemisphere ridge anomaly) and their sum (red), from January of the precursor year through December of the event year. "Ridge" refers throughout to the climatological subtropical high itself; SLP_R2a and SLP_R5a are its anomaly, and a positive correlation with A2 means the ridge strengthens when A2 is positive and weakens, in the direction that favors El Niño, when A2 is negative. The three labelled maxima are peaks in this correlation profile, not episodes of ridge intensification. Yellow shading marks the A2 sampling window; dashed lines mark the phase-randomized Monte Carlo 95% threshold (|r| = 0.27). (b) Correlation of A2 with sea level pressure in Y−1 August, the end of austral winter: within one month of the closure of the A2 window the Southern Hemisphere ridge anomaly is already significantly correlated with A2 (SLP_R5a, r = +0.37), while the Northern Hemisphere ridge is not (SLP_R2a, r = +0.19). At this stage the significant region sits mostly west of the dateline at midlatitudes in each hemisphere, connected across the equator only by a narrow and broken corridor, and the R2 box lies at its edge rather than within it; read against the framework, this is an early and geographically incomplete engagement, the seed of the sequence rather than a configuration that on its own favors ENSO onset. (c) Y−1 December: the significant region has consolidated onto the eastern side of the basin,

now centered near the R2/R5 longitudes, and forms one coherent structure spanning the equator in which both ridges and the corridor between them are simultaneously significant (SLP_R2+R5a, r = +0.41), with 86% of the 30°S–30°N, 210–240°E sector exceeding the threshold. (d) Y0 May: the coherent structure has been carried poleward, reaching about 40°N while its southern flank retreats to about 22°S, and now solidly overlies the Northern Hemisphere ridge, which dominates (SLP_R2a, r = +0.38). Stippling and the heavy contour mark $|r| > 0.27$; thin contours are drawn at 0.15, 0.30 and 0.45. Green boxes: R2 (20–30°N) and R5 (20–30°S), 210–240°E. Between August and May the significant region moves from a western, equator-sparse configuration to a coherent eastern structure spanning the equator and finally reaching poleward over the ridges themselves, so that the subtropical highs are engaged three times in succession rather than once.

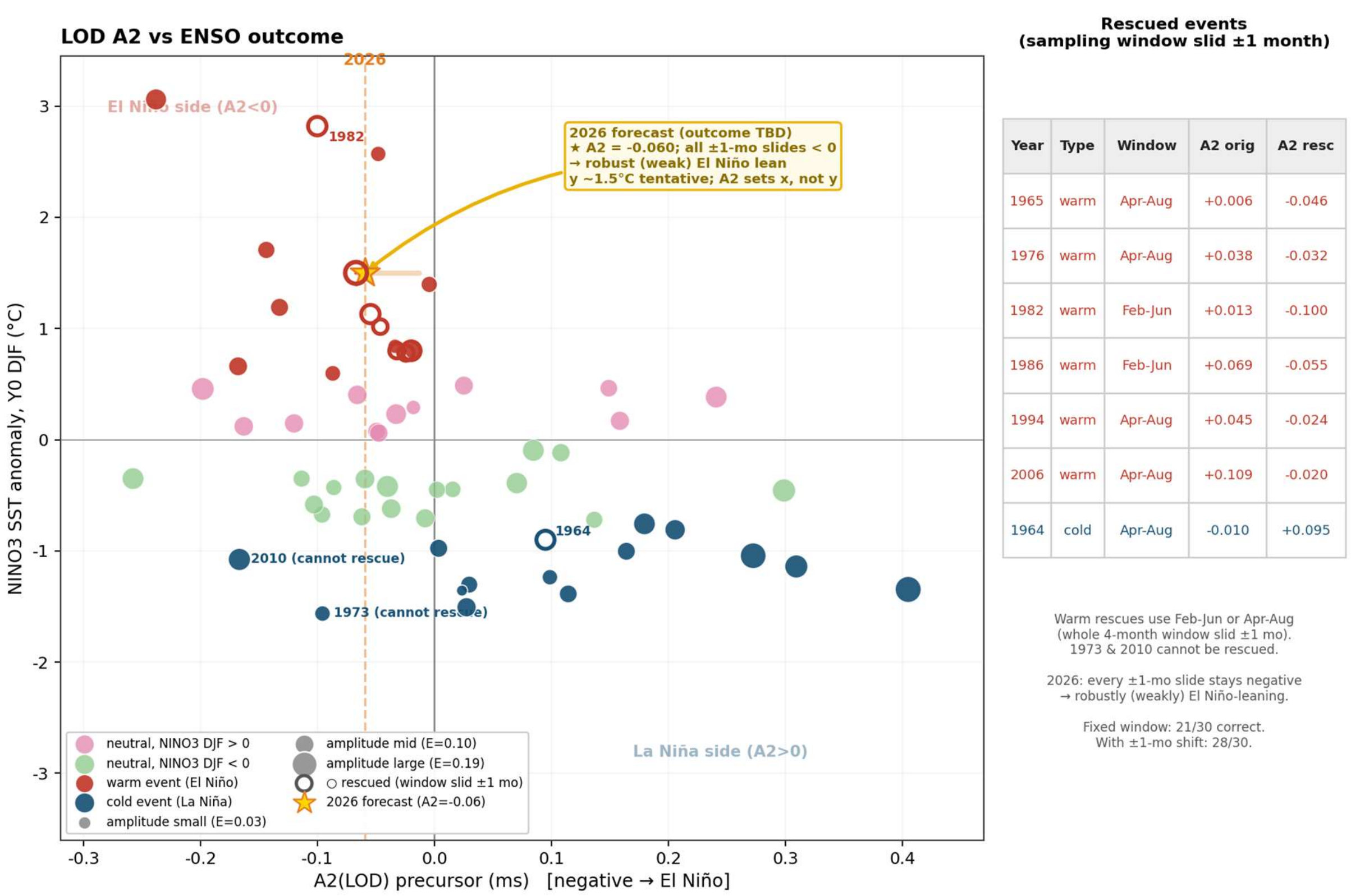

| Year | Type | Window | A2 orig | A2 resc |
|---|---|---|---|---|
| 1965 | warm | Apr-Aug | +0.006 | -0.046 |
| 1976 | warm | Apr-Aug | +0.038 | -0.032 |
| 1982 | warm | Feb-Jun | +0.013 | -0.100 |
| 1986 | warm | Feb-Jun | +0.069 | -0.055 |
| 1994 | warm | Apr-Aug | +0.045 | -0.024 |
| 2006 | warm | Apr-Aug | +0.109 | -0.020 |
| 1964 | cold | Apr-Aug | -0.010 | +0.095 |

FIG. 9. End to end: the A2 precursor against the ENSO outcome, and the role of phase drift. Horizontal axis: A2 from the fixed March-minus-July window of the precursor year; vertical axis: NINO3 sea surface temperature anomaly in the following DJF. Marker area is proportional to the amplitude of the sub-annual oscillation in the precursor year, so a large symbol at small |A2| identifies a year in which the oscillation was strong but its phase fell between the sampling months. Red and blue filled circles are El Niño and La Niña events; pink and green circles are neutral years, separated by the sign of the observed anomaly. Across the fixed window the correlation is $r = -0.33$ ($P = 0.011$) and 21 of the 30 events are correctly signed. Open circles show the seven events that a one-month shift of the whole four-month sampling window moves to the correct side; the accompanying table lists each year, the shifted window, and the original and shifted values of A2. With the shift, 28 of 30 events are correctly signed, including every El Niño. Two cold events, 1973 and 2010, are not recovered by any admissible shift. The dominant period of the oscillation is about eight months (median crest-to-crest spacing of 8 months over 89 cycles; the supplemental

material), and its crests fall in every calendar month with near-uniform frequency, so its extrema wander relative to any fixed calendar window; the one-month shifts therefore measure phase drift in the oscillation rather than tune a forecast, and they are applied here retrospectively, with knowledge of the outcome. The gold star marks the 2026 forecast (A2 = −0.060 ms). All nine one-month window choices remain negative (−0.067 to −0.013), so the phase indication leans El Niño and is robust to phase drift, while the small amplitude places the year in the lowest confidence tercile: the phase is the reliable part of the message, and the vertical position of the star is provisional pending the observed outcome.

Beyond single correlations, formal mediation analysis (Fig. 10; full table in supplemental Table S2) establishes the statistical structure of the chain through partial correlations alone, with no resampling required: the total effect of A2 on the ocean is significant, the mediator is engaged and transmits, and the direct residual collapses to insignificance once any candidate mediator is held fixed. A significant total effect combined with a vanishing direct residual is the signature of complete mediation, and the two facts are not in tension: the uncontrolled correlation mixes the mediated and any unmediated contribution together, so once the mediated portion is removed, a genuinely vanishing residual is exactly what should remain if there is no unmediated contribution left. Here that is what is observed: everything A2 delivers to the ocean travels through the ridges and the equatorial wind, a verdict a bootstrap product-of-coefficients analysis (5,000 resamples) reaches independently in every configuration (supplemental material).

Two further checks address the direction-of-influence question directly. First, the band carries essentially no amplitude-coherent ENSO echo (maximum cross-correlation 0.05, against 0.40 in the classical 2 to 10 yr response band), an orthogonality intrinsic to the band rather than a product of harmonic removal (section 7b; supplemental material). Second, the chain does not reduce to ENSO persistence: with the preceding winter's NINO3 partialled out, the December and May ridge links and the June wind link all survive, several strengthening (supplemental material). The EPSH and wind links are therefore the oscillation's own content rather than an inherited signature of the previous event; the sequence by which they arrive is examined in the discussion.

**The EPSH is the conduit: the only route from Earth rotation to the ocean runs through the ridges**

−0.14 to −0.20 (P = 0.14 to 0.31) not significant: nothing bypasses the EPSH

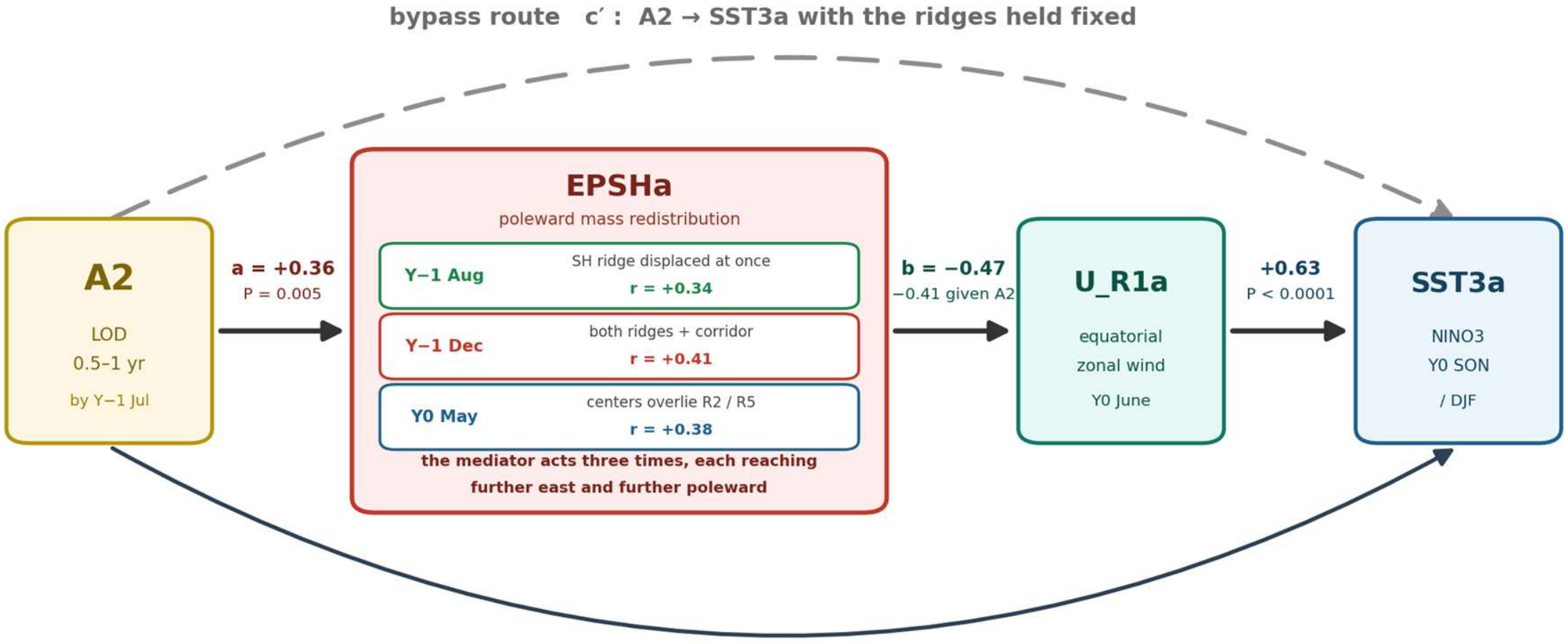

**total effect c : A2 → SST3a with nothing held fixed = −0.33 (P = 0.010) significant**

The total effect is real. It is carried entirely through the ridges, which is why the bypass above is empty.

**Partial correlations, no resampling required (n = 59 years)**

| what is held fixed | A2 → SST3a (SON) | A2 → SST3a (DJF) | EPSH → SST3a \| A2 |
|---|---|---|---|
| nothing: this is c | −0.31 (P = 0.017) | −0.33 (P = 0.010) | −0.53 |
| SLP_R2a (Y0 May): c′ | −0.14 (P = 0.31) | −0.16 (P = 0.22) | — |
| SLP_R2+R5a (Y0 May): c′ | −0.15 (P = 0.25) | −0.18 (P = 0.18) | −0.47 (P = 0.0002) |
| U_R1a (Y0 June): c′ | −0.17 (P = 0.21) | −0.20 (P = 0.14) | — |

Row 1 is the total effect c. Rows 2 to 4 are the direct residual c′, what remains of A2 once the mediator is held fixed; none of them is significant.
The last column is path b: the mediator still commands the ocean after A2 is removed. The EPSH to U_R1a links established earlier in this paper
(SLP_R2a in February, r = −0.59; SLP_R5a in September, r = −0.77; both $P < 10^{-5}$) corroborate path b independently. Bootstrap resampling agrees (appendix C).

FIG. 10. The EPSH is the conduit from Earth rotation to the ocean (n = 59 years, 1964–2022). The mediator is not an instantaneous relay: within the EPSH box the three stages of Fig.8 are shown, the southern ridge displaced in Y−1 August (r = +0.37), both ridges and the corridor between them in Y−1 December (r = +0.41), and the significant centers overlying the ridges themselves in Y0 May (r = +0.38), each redistribution reaching further east and further poleward than the last. Standardised path coefficients are given on the straight arrows, the value in parentheses on path b being the partial correlation after A2 is controlled. Two arcs span the chain and must not be confused. The solid arc below is the total effect c, the plain correlation between A2 and SST3a with nothing held fixed; it is significant (c = −0.33, P = 0.010). The dashed arc above is the direct residual c′, what is left of that effect once the ridges are held fixed; it is not significant under any choice of control variable (c′ between −0.14 and −0.20, P = 0.14 to 0.31). A significant total effect together with a vanishing direct residual is the definition of complete mediation: the effect of Earth rotation on the ocean is real, and it travels entirely through the subtropical ridges. Mediation is established here by partial correlation alone, with no resampling. The inline table condenses the full mediation table (supplemental material, Table S2): its first row is c, its remaining rows are c′ under each mediator, and its last column is path b, showing that the mediator still commands the ocean after A2 is removed (r = −0.47, P = 0.0002). The EPSH to U_R1a relationships established earlier in this paper (SLP_R2a in February, r = −0.59; SLP_R5a in September, r = −0.77; both $P < 10^{-5}$) corroborate path b independently of the sub-annual analysis. Bootstrap resampling (5,000 draws) reaches the same verdict and is reported in the supplemental material.

### *d. The Y+1 LOD reversal: geodetic fingerprint of event termination*

Extending the composite window from 24 to 36 months reveals a systematic Y+1 LOD phase reversal, the planetary-scale geodetic fingerprint of the MSIGMA termination mechanism that we describe dynamically in the next section. The extended 36-month composite figure (Fig. E1 of the supplemental material) shows that El Niño events associated with a positive Y−1 LOD phase transition to a negative Y+1 LOD phase, and vice versa for La Niña. This opposite-sign structure of the Y−1 precursor and Y+1 response also explains why earlier undifferentiated LOD–ENSO correlation studies found weak net signals (Zheng et al. 2003): the two contributions partially cancel when all temporal lags are combined without explicit temporal window separation. By contrast, the 2–10 yr band maintains strong El Niño/La Niña separation throughout Y0 (Fig.6e), consistent with this band reflecting the concurrent ENSO response rather than the upstream precursor.

## 6. A unified hierarchical framework

The results can be synthesized into a four-tier hierarchical framework (Fig.11) characterized by three features: two silent orchestrators working behind the scenes (Earth's rotation and the annual seasonal cycle), a spatial scope spanning the mid-latitude Pacific to the tropics with planetary-scale angular momentum exchange, and an action pathway from solid Earth through the atmosphere to the tropical ocean. The framework portrays ENSO as a complex process in which the "script" is set by low-frequency, deterministic external forcings, and "triggered" at critical moments by higher-frequency, stochastic internal forcings.

### *a. Tier 1: the annual cycle as the metronome*

The Earth's revolution and the tilt of its axis of rotation cause the change of seasons, which is the most powerful and regular external forcing in the climate system. The annual cycle determines the seasonal windows in which each subsequent tier operates: the NH EPSH trigger phase in boreal winter–spring, the SH EPSH sustainer phase in boreal summer–autumn, and ENSO phase-locking toward year-end (Stuecker et al. 2013). The annual cycle is the metronome orchestrating the hemispheric handoff but does not itself initiate events.

### *b. Tier 2: LOD/AAM and MJO stochastic triggers*

LOD and AAM are physically inseparable: by angular momentum conservation, an increase in atmospheric angular momentum must be compensated by an equal LOD increase (Rosen and Salstein 1983). The LOD (0.5–1 yr) oscillation identified as an ENSO precursor operates through this AAM pathway. The MJO, on 30–60-day intraseasonal timescales, drives sub-seasonal LOD/AAM variations (Weickmann et al. 1997) and excites westerly wind bursts that can initiate El Niño warming (McPhaden 1999; Kessler 2001). Conversely, the present study identifies a sub-annual LOD signal that engages the ridges in three successive stages and preconditions the coming event fourteen to seventeen months before its peak. The ultimate attribution of this signal should not be over-concluded in either direction. In the 0.5 to 1 yr band LOD tracks atmospheric angular momentum at $r = 0.97$, so the proximate content of the signal is atmospheric, shaped by the seasonal-scale interplay of SST, tropical convection, the jets and the trade winds (Egger et al. 2007). At the same time, the angular momentum of the atmosphere is exchanged with the solid Earth through friction and mountain torques acting across the planetary topography, so the solid Earth participates in the exchange rather than merely recording it. Establishing which torque channel carries the sub-annual signal, and how well the angular-momentum budget closes on these seasonal timescales, requires a dedicated torque analysis that lies beyond the scope of this paper. We therefore leave the attribution open. LOD serves as the integrated, reanalysis-independent read-out of this angular-momentum state; its value lies in providing a single continuous geodetic observable that captures the upstream atmospheric precursor.

Whether the sub-annual LOD oscillation predisposes the system toward enhanced MJO activity, or whether both respond independently to a common angular momentum source, cannot be resolved from the present analysis. Both LOD/AAM and the MJO are therefore placed in Tier 2 as related stochastic forcings that operate at different timescales: the sub-annual LOD oscillation as a probabilistic preconditioner months in advance, and the MJO as the final intraseasonal trigger once the system is preconditioned by Tier 2a and Tier 3.

### *c. Tiers 3 and 4: the EPSH pacemaker and equatorial wind forcing*

Low-frequency EPSH anomalies, modulated by the annual cycle, constitute the primary pacemaker (Tier 3): NH EPSHa acts as the trigger, while SH EPSHa acts as the sustained amplifier; their synchrony or asynchrony determines event type and magnitude. The EPSH acts through U_R1a (Tier 4), which deepens the eastern Pacific thermocline through forced Kelvin waves and initiates the Bjerknes positive feedback; it is at this tier that the anomaly, once it reaches the ocean, is amplified and given the memory that carries an event through its full life cycle.

### *d. Termination: the Mirror-SIGMA (MSIGMA) pattern*

At ENSO maturity, anomalous tropical heating generates a coupled Hadley–Walker circulation response creating the Mirror-SIGMA (MSIGMA) sea level pressure anomaly pattern (Wang and Qin 2023). This pattern exhibits paired positive SLPa bulges near the subtropical dateline in both hemispheres, produced by intensified subtropical subsidence associated with the anomalous meridional cell, together with an equatorial negative SLPa wedge intruding westward from the eastern Pacific, reflecting the weakened Walker circulation. The boundary line between these two opposing SLP anomaly regions resembles a mirror-reversed Greek letter "Σ," hence the name. These MSIGMA bulges strengthen the EPSH, reversing the El Niño-driving forcing and preconditioning the basin for La Niña.

Notably, the upstream role of LOD(0.5–1 yr) as an ENSO preconditioner diminishes by Y0 mid-year, as the cold/warm composite divergence in this band begins to converge (Figs.6b and 8a). By contrast, the 2–10 yr band maintains strong El Niño/La Niña separation throughout Y0 (Fig.6e), consistent with this band reflecting the concurrent ENSO response rather than the upstream precursor. The 36-month composite (Fig. S3) reveals the full temporal succession: from LOD as preconditioner (Y−1), through MSIGMA as terminator (Y0 autumn onward), to LOD as response (Y+1), when the systematic phase reversal in the 0.5–1 yr band reflects the downstream ENSO signal through AAM redistribution rather than a continuation of the upstream mechanism. This temporal succession closes the solid Earth–atmosphere–ocean loop and provides independent geodetic confirmation that the ENSO cycle has entered its decay phase.

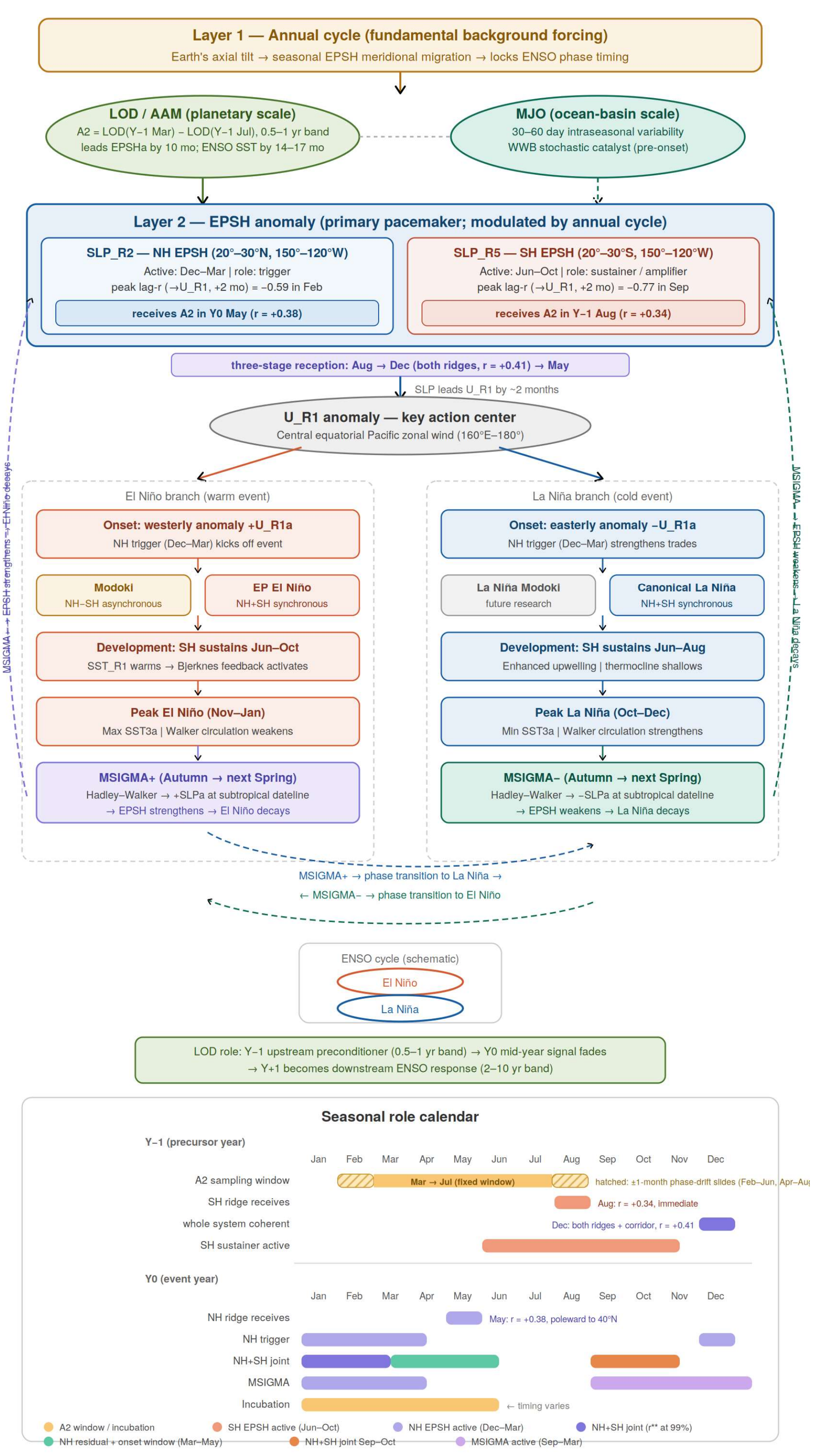

FIG. 11. Unified hierarchical framework for ENSO. The diagram presents the complete ENSO lifecycle as a four-tier forcing hierarchy with MSIGMA feedback closure. Tier 1 (Annual cycle):

seasonal EPSH migration of approximately 5° latitude northward in boreal summer sets the trigger, sustainer, and phase-locking windows. Tier 2 comprises two stochastic forcings at different timescales: Tier 2a (LOD/AAM) is a probabilistic preconditioner: its sub-annual oscillation (dominant period about eight months) engages the ridges in three successive stages that together precondition the coming event fourteen to seventeen months before its peak. Tier 2b (MJO) provides the final intraseasonal trigger through 30–60-day westerly wind bursts. Tier 3 (EPSHa, primary pacemaker): the NH EPSH (SLP_R2a) acts as the trigger (active December–March, peak 2-month lead $r = -0.59$ in February); the SH EPSH (SLP_R5a) acts as the sustainer (active June–October, peak $r = -0.77$ in September). Tier 4 (U_R1a): central equatorial Pacific zonal wind at 160–180°E, the relay through which the ridge anomaly reaches the ocean and engages the Bjerknes feedback. Below the tier hierarchy, the El Niño branch (left) and La Niña branch (right) each proceed through four phases: onset (with a diversity fork distinguishing Modoki from canonical events), development, peak maturity, and termination via MSIGMA. Dashed side arrows show MSIGMA feedback closing the cycle. The LOD role evolution box summarizes the temporal succession from upstream preconditioner (Y−1) through fading signal (Y0 mid-year) to downstream ENSO response (Y+1).

## 7. Discussion

### *a. Resolving the four classical puzzles*

The hierarchical framework resolves all four classical ENSO puzzles simultaneously. The trigger problem yields to a multi-tier cascade: the re-phased sub-annual angular-momentum oscillation, read out by LOD, preconditions the subtropical highs through their seasonal receptive windows in three successive stages and thereby alters the prior probability of subsequent ENSO development, NH EPSHa provides the deterministic seasonal trigger, and MJO wind bursts deliver the final stochastic impulse. Because no single factor is both necessary and sufficient, the framework naturally accounts for the observed irregularity of ENSO onset. Cycle irregularity likewise emerges from variable EPSH synchrony and persistence, generating the observed 2–7 year period range without requiring nonlinear chaos in the tropical ocean alone. The diversity problem is resolved by the synchrony distinction demonstrated in the 1991/92 and 1997/98 contrasting cases. The spring predictability barrier arises because the critical NH-to-SH handoff occurs when the inter-hemispheric EPSHa correlation is not significant (Table 3); monitoring SH EPSHa and the preceding-year LOD amplitude offers a physically grounded path beyond this barrier.

### *b. A sub-annual angular-momentum rhythm, gated by the seasonal subtropical highs*

The individual correlations in this chain are modest, and it is worth saying at the outset why that is the signature the proposed mechanism should leave rather than an argument against it. Length of day is not the actor here; it is the instrument. In this band it tracks the atmosphere's axial angular momentum almost perfectly ($r = 0.97$ with AAM), so a length-of-day fluctuation is the solid Earth's faithful record of a sub-annual swing in atmospheric angular momentum. Consistent with this, repeating the correlation-field analysis with A2 defined on atmospheric angular momentum in place of length of day yields fields indistinguishable from those of Fig.8, confirming that the geodetic series is read here purely as an atmospheric observable. Repeating the construction of A2, and of the correlation fields of Fig.8, with reanalysis AAM in place of LOD reproduces both, so the geodetic and the reanalysis readings of this band are interchangeable. Such a swing cannot reach the equatorial ocean directly. It reaches it by rearranging mass, and the eastern-Pacific subtropical highs are where that rearrangement becomes dynamically consequential, because it is there that the ridges govern the trade winds feeding the central equatorial Pacific. The subtropical high is therefore the protagonist of this story, and length of day is the geodetic instrument that lets us watch the swing arrive.

It is worth setting out the angular-momentum bookkeeping explicitly, because it fixes every sign in the chain and because the same conservation law has long been read in the opposite direction. The atmosphere and the solid Earth exchange axial angular momentum only through friction and mountain torques (Barnes et al. 1983; Egger et al. 2007), and their sum is conserved on the timescales considered here. When length of day increases the solid Earth is turning more slowly, having given up westerly angular momentum to the atmosphere. An atmosphere richer in westerly angular momentum is one in which the tropical easterlies, the trade winds, are weaker; and since the trades are the low-level outflow of the subtropical highs, weaker trades accompany weaker ridges. The subtropical high, the trade winds and the rotation of the Earth are thus three readings of one quantity, and the reverse holds for a shortening day. Every sign in the observed chain, from A2 through the three ridge stages to the equatorial wind and the ocean, obeys exactly what this bookkeeping requires; none has to be assumed, and the itemized verification is given in the supplemental material.

A further point resolves what might otherwise appear to be a paradox of timescales. Length of day and atmospheric angular momentum vary together essentially without delay ($r = 0.97$ in this band), and the sub-annual swing itself completes within months; yet its consequences remain legible in the ridges ten months after the sampling window closes. The reconciliation is that the swing writes an imprint on the atmospheric mass field that outlives the swing itself. The first entry is immediate:

within one month of the window closing, the poleward momentum transport accompanying the decaying swing registers as the August maximum of the SLP_R2+R5a lead profile (Fig.8a), carried at that stage almost entirely by the Southern Hemisphere ridge (Fig.8b). What persists thereafter is not the oscillation but its record, renewed and extended by the oscillation's regular propagation: each subsequent passage deposits its anomaly on ground prepared by the last, so that the significant region spreads eastward and poleward across the eastern Pacific ridge domain in successive overlapping advances, producing the December and May maxima in turn (Figs.8c,d). The atmosphere, on this reading, is not a passive conduit that forgets each swing as it passes but a recording medium on which the swing leaves a cumulative mark. Only at the end of this progression does the imprint acquire its influence on the ocean: the fully engaged ridges command U_R1a, and U_R1a commands SST3a, the two links established independently in the first half of this paper, and it is through them that the geodetic signal is converted into a systematic influence on ENSO development.

The same physics has long been read in the reverse direction, and it is important to be clear about why we may read it forward here. During an El Niño the trades collapse, the atmosphere gains angular momentum and the day lengthens, so length of day is conventionally, and correctly, understood as a consequence of ENSO rather than a cause (Chao 1988; Dickey et al. 1994), with the observed one to two month lag explained dynamically by thermal winds arising from the poleward gradient of tropical temperature (Dickey et al. 2007). Conservation of angular momentum constrains the partition between atmosphere and solid Earth; it does not, by itself, assign a direction of causality. The direction must be established by other means, and the means here is the frequency separation reported above: the classical response occupies the 2 to 10 yr band, while the 0.5 to 1 yr band carries no amplitude echo of ENSO at any lag. Whatever generates the sub-annual fluctuation, the ocean does not; an anomaly that ENSO did not create, but which nonetheless rearranges atmospheric mass across the subtropical ridges, can only be read as acting upon the system it precedes.

This bookkeeping suggests a simple energetic reading of why a rise in length of day should precede a warm event, which we advance as a working hypothesis. The easterly trades owe their direction to Earth's rotation: they are the Coriolis-deflected equatorward branch of the Hadley circulation, organized at low levels by the subtropical highs. Blowing steadily on the equatorial Pacific, they bank warm surface water in the west, depress the thermocline there and lift it in the east, and hold the ocean in a tilted state whose zonal pressure gradient balances the wind stress (Bjerknes 1969). The tilt is stored energy: the trades wind the spring, and rotation, through the Coriolis force, sets the direction in which it is wound. A rise in length of day across the March to July window marks a transfer of westerly angular momentum from the solid Earth to the atmosphere, that is, a systematic weakening of the easterlies at precisely the season when the sub-annual oscillation is phase-locked to the annual march of the ridges. A weakened trade system can no longer support the accumulated gradient, and the stored potential energy discharges eastward, as warm-pool water surges east and equatorial Kelvin waves flatten the thermocline. This is in essence the relaxation mechanism proposed by Wyrtki (1975), here supplied with an extratropical, rotational timer. In this reading the sub-annual angular-momentum swing does not create the El Niño; it unlocks a release that rotation has already paid for.

The energetic core of this picture deserves to be stated in its own terms. The east to west tilt of the equatorial Pacific, the sea surface standing high in the west and the thermocline deep beneath the warm pool, is not an accumulation that each cycle must first build; it is the standing configuration that rotation maintains, through the trades, for as long as the planet turns. The permanence of the warm pool is the evidence: its surface temperature varies by only a fraction of the eastern equatorial variability, because the reservoir behind it is never allowed to empty. The reservoir breathes, drawing down through an event and refilling afterward, but under the standing forcing it is rarely far from ready; full or half full, it can discharge if the window is struck. What is scarce is therefore not energy but timing. Release occurs when the sub-annual upswing of length of day lands inside the seasonal window, and in the years it does not, nothing is discharged and the system simply holds its charge. Occurrence is thus gated by a coincidence, random in any given year yet certain to recur, between an oscillation of roughly eight months that wanders freely through the calendar and an annual gate that does not move. We suggest this as a natural origin for the irregularity of ENSO: the timer sets the phase of each event, the state of the reservoir at the moment of release sets its amplitude, and the statistics of the coincidence set the irregular intervals between events.

The correlation fields show a single process unfolding, not a set of isolated coincidences (Fig. 8). Within one month of the closure of the A2 window the southern ridge has already answered, while the northern ridge remains inactive; nothing propagates slowly here. An angular-momentum anomaly of the whole atmosphere is, by construction, a rearrangement of atmospheric mass, and the southern high responds where and when it is most easily displaced, at the end of austral winter when that ridge is at its strongest. Through the months that follow, the anomaly moves both poleward and eastward: by December the whole eastern-Pacific system is engaged at once, both ridges together with the corridor between them, and by the following May the significant region has been carried poleward in each hemisphere until its centers lie squarely over the R2 and R5 boxes themselves. Three times, then, the oscillation drives mass poleward and three times the subtropical highs are displaced, each occasion reaching further east and further poleward than the last. The third is the consequential one, not because its correlation is the largest but because by May the significant centers coincide with the ridges themselves, where pressure anomalies are largest and where, as established earlier in this paper, the EPSH governs U_R1a and through it SST3a. Significance and influence are not the same quantity.

Two facts explain why the sub-annual band, and no other, is the place to read this sequence, and why the signal is not the previous ENSO event returning in disguise. The first is that this band is almost blind to ENSO's own amplitude: the classical ENSO-driven response of Earth rotation lives in the 2 to 10 yr band, while the cross-correlation of the sub-annual band with NINO3 never exceeds 0.05 at any lag, an orthogonality that holds independently of whether the seasonal harmonics are removed (supplemental material). The second fact is subtler: the preceding event does organize this band, but through phase rather than amplitude, correlating strongly with the previous winter's NINO3 while sharing almost no amplitude coherence with it; we read this as the termination of an event resetting the phase of the oscillation without feeding energy into it, a hand-off plausibly tied to the Mirror-SIGMA reorganization that closes the cycle (section 6d), and confirmed by the fact that the downstream chain survives essentially undiminished once the previous winter's NINO3 is held fixed (supplemental material).

What this buys is not a forecast index competing with existing ones but a physical account with a long observational reach. The clearest evidence is structural rather than statistical: setting the timeline by the 30 strong events of the record and asking only what the sub-annual oscillation does before each of them, the composite evolution of length of day is a mirror image between warm and cold events, the same rhythm and the same seasonal timing with every sign reversed. An oscillation that merely echoed the ocean could not produce this symmetry ahead of the events that define it. Because A2 requires the July value of the precursor year, the index is complete some fourteen to seventeen months before the event reaches its peak. Within that reach it fixes the phase of the coming event, warm or cold, and not its amplitude, and it does so only when the swing lands inside the window, exactly as Fig. 9 shows event by event. The small fraction of misses is concentrated in years when the oscillation's roughly eight-month period placed its crest outside the sampling window; the case-by-case rescue statistics behind this figure are given in the supplemental material.

### *c. A two-stage prediction strategy*

The framework suggests a two-stage prediction strategy. Stage 1 (LOD-based extratropical preconditioning, complete by the end of July of the precursor year, fourteen to seventeen months ahead of the ENSO SST response): the LOD(0.5–1 yr) oscillation amplitude and phase shift the prior probability of subsequent ENSO development of the corresponding sign, a signal independent of SST predictors and available before the spring barrier. Stage 2 (3–6 months ahead): the synchrony of NH and SH EPSHa in boreal spring discriminates event type, complementing classical subsurface heat-content and SSTa monitoring (Lopez and Kirtman 2014; Clarke and Zhang 2019).

The precursor is best used as a confidence-stratified warning rather than a uniform predictor. Because A2 is the product of the oscillation amplitude and its phase projection onto the fixed sampling window, |A2| itself measures how completely the signal is in place: the top tercile of |A2| shows a 75% hit rate ($r = -0.61$, $P = 0.005$), while the lower two terciles are indistinguishable from chance. Small |A2| therefore marks a genuinely silent state rather than a failed forecast; a large-|A2| year is an actionable warning, a small-|A2| year an honest abstention.

Two further points sharpen the practical reading. First, the fixed sampling window is conservative with respect to phase drift, as Fig. 9 shows event by event; a selection-aware significance analysis, charging the Monte Carlo surrogates for the same window search, is reported in the supplemental material. Second, as a live test, the current IERS series gives A2(2026) = −0.06 ms, negative for every one of the nine ±1-month window choices, so the phase indication for 2026 leans toward El Niño and is robust to phase drift, while its small amplitude places it in the low-confidence tercile: the phase is the stable part of the message, the intensity is not.

The 1997/98 super El Niño illustrates this prediction potential concretely (Fig. S4). With A2 ≈ −0.24 ms, among the largest negative values in the 59-year record, the LOD amplitude signal was available by mid-1996; roughly 10 months before the EPSH reached peak negative anomaly (May 1997) and 17 months before the SST peak, well ahead of the boreal spring 1997 predictability barrier.

***d. Two metaphors: the cart and the sluice gate***

To fix the physical logic in a single image, picture the ENSO system as a cart drawn by two horses whose gait is neither steady nor uniform but quickens and slackens in rhythm with the seasons. The horses are the two subtropical highs, a stronger, steadier southern horse paired with a weaker, more variable northern one, and the traces through which they pull are the trade winds acting on U_R1a. When the team quickens, the traces tighten and the cart is hauled forward: a transient acceleration of Earth's rotation, arriving in the matching seasonal phase, strengthens the subtropical highs, firms the trades, and tilts the system toward La Niña. When the team slackens, the traces go loose while the cart coasts on under its own momentum: a transient deceleration, arriving in the opposite phase, weakens the highs, relaxes the trades, and lets the warm water the trades had banked in the west spread eastward, engaging the Bjerknes feedback and favoring El Niño. What decides whether an event is initiated is not any steady quickening or slowing but the alignment of these rhythmic surges with the seasonal march of the highs, the moment in the stride at which the pull arrives. And whether the outcome is a canonical event or a Modoki depends on whether the two horses change gait together or one leads while the other lags.

A second image, less kinetic and more hydraulic, may fix the same logic from another angle. Picture the warm pool as a standing reservoir, kept full by a permanent works: the trades, which owe their existence to Earth's rotation, wall the water in the west day after day, event or no event, and the reservoir is rarely far from capacity because the wall never comes down between one El Niño and the next. On this reading the ridges are the sluice gate set into that wall, and the sub-annual rotation oscillation is the keeper who happens to pass the gate on its rounds. For most of the year the keeper passes and nothing happens, because the gate is elsewhere in its own seasonal cycle and the two do not meet. It is only when the keeper's rounds bring it to the gate at the moment the annual cycle has already swung the gate toward open, the March to July window engaging the ridges through their own late-spring stage, that the keeper's push is enough to swing it the rest of the way, and the standing water is released eastward as El Niño. On this reading rotation neither fills the reservoir nor builds the wall; it is the rotating keeper who decides, by the accident of timing, whether this particular round finds the gate ready to give.

### *e. Limitations and open questions*

The analysis rests on 59 years of observations (approximately 9–10 statistically independent ENSO cycles), and we do not claim that every result is beyond further scrutiny. Five independent and mutually reinforcing lines of evidence make the core conclusions robust (the supplemental material): spectral composites across 30 ENSO events; full-sample lead–lag correlations; two sequential mediation analyses; a sensitivity check (supplemental Table S3 of the supplemental material) showing that all three chain correlations remain significant at $P < 0.01$ after removing the three strongest El Niño events; and the well-documented 1997/98 case study with correct temporal ordering at 2–3 month resolution. Two limitations warrant explicit acknowledgement. First, the analysis is entirely observational and awaits validation in coupled climate model experiments, especially, targeted experiments with prescribed LOD perturbations is necessary. Second, stationarity of the LOD–EPSH relationship is assumed; although the underlying mechanism depends on the phase alignment between two physically determined periodicities, multi-decadal modulation by the Pacific Decadal Oscillation or anthropogenic warming cannot be excluded without longer records or model projections.

## 8. Conclusions

This study has proposed a new dynamical theory for ENSO driven by the EPSH, with Earth rotation variability as an upstream preconditioner and the annual cycle as an overall metronome. The core idea is that the ENSO cycle is not an isolated tropical phenomenon but a Pacific basin-wide, air–sea–solid-Earth coupled process profoundly constrained by the seasonal evolution of extratropical atmospheric circulation and angular momentum exchange. Several key conclusions can be drawn from the evidence presented.

A methodological result underpins the whole chain and deserves separate statement. It has long been recognized that ENSO leaves its mark on Earth's rotation through the annual cycle: the amplitude of the annual harmonic of length of day, fit year by year, covaries with that same year's boreal winter NINO3 at $r = -0.31$ ($P = 0.015$), the response direction long documented in the literature (Chao 1988; Dickey et al. 1994). That harmonic sits directly on the edge of the sub-annual band and, left in place, would inflate it threefold; removing the annual and semi-annual terms deterministically excises this route (derivation in the supplemental material). What then remains in the sub-annual band carries no amplitude echo of ENSO whatever, its cross-correlation with NINO3 never exceeding 0.05 at any lag, while the same record retains the classical ENSO-driven response at 0.40 in the 2 to 10 yr band. Two signals of opposite causal direction therefore occupy the same geodetic record in separate frequency bands, and isolating the sub-annual one is what permits Earth rotation to be read as a precursor rather than as a consequence. To our knowledge this separation has not previously been demonstrated, and it is the license on which the rest of this work rests.

First, the fundamental driving force of the ENSO cycle originates from anomalies of the NH and SH EPSH, while the annual cycle, through its seasonal modulation of the EPSH, acts as the metronome that locks the timing and development rhythm of ENSO events. Second, the NH EPSH anomaly in winter and spring typically acts as the trigger, while the SH EPSH anomaly in summer and autumn plays the role of sustainer and amplifier; the degree to which these two hemispheric forcings synchronize or differ in phase largely regulates the intensity, pattern, and type of ENSO events, including EP and CP/Modoki events. Third, the initial response action center of ENSO to extratropical forcing is located in the central equatorial Pacific near the dateline; the MSIGMA SLP pattern formed during the mature phase, by exciting an anomalous regional Hadley circulation, feeds back to the subtropical high and is the key mechanism facilitating ENSO phase transition. Fourth, a stable sub-annual precursor signal exists in the LOD (0.5–1 yr) spectral band, and when its phase aligns with the annual EPSH rhythm it engages the ridges in three successive stages, in Y−1 August, Y−1 December, and Y0 May, the significant correlation field expanding eastward and poleward at each stage until, by Y0 May, it fully overlies the SLP_R2 and SLP_R5 boxes themselves. This progressive engagement is the atmospheric imprint of a swing that is itself brief, the mechanism by which a signal in the solid Earth comes to precondition the ENSO cycle some fourteen to seventeen months before its peak. LOD is thus an active upstream participant, not merely a passive downstream responder.

Fifth, this precursor is not a restatement of the classical rotation-ENSO connection but its counterpart in a separate frequency band: the annual harmonic of length of day carries the long-recognized concurrent response of rotation to a mature event, while the sub-annual band used throughout this paper is orthogonal to ENSO at every lag and is read here in the forward direction alone. Mediation analysis by partial correlation further shows that the ridges are not merely correlated with the precursor but are its sole conduit to the ocean, the direct path collapsing to insignificance once the ridges are held fixed. Taken together, the frequency separation and the mediation result convert what has long been treated as a downstream curiosity of Earth rotation into an upstream constraint on when, within the ENSO cycle, an event becomes likely.

These findings have practical implications for ENSO prediction. Traditional forecasting relies on subsurface heat content and tropical SSTa monitoring; the new theory indicates that the true root extends farther upstream, to the extratropical atmosphere and to solid-Earth rotation variability. A more effective strategy would incorporate the lagged SLP–U_R1 relationship (Fig. 2) and the A2-type LOD amplitude index into existing ensemble systems. Future work should validate these conclusions in coupled climate models and develop practical prediction tools incorporating EPSH and upstream LOD/AAM indicators. Reframing the ENSO pacemaker as an extratropical–tropical coupled system embedded within a solid-Earth–atmosphere–ocean chain opens new avenues for diagnosis, theoretical interpretation, and prediction on global scales.

## Acknowledgements

The author thanks Professor K. Yamazaki for invaluable academic advice and Professor Y. Fujiyoshi for ongoing support.

The author would like to express his special gratitude to Dr. Klaus Weickmann for his thorough review of the manuscript and for numerous constructive comments that identified gaps in the original analysis and interpretation. The extended discussions that followed were both scientifically productive and personally instructive, and they led to significant improvements in the clarity and rigor of this work. His generosity with time and insight is sincerely appreciated.

## Author contributions

Y. Wang conceived this research and completed all the work on this paper, including producing the figures and tables, analyzing the data and writing the manuscript.

## Competing interests

The author declares no competing interests.

## Data availability statement

NCEP/NCAR reanalysis data are available from the NOAA Physical Sciences Laboratory (https://psl.noaa.gov/). NOAA Extended Reconstructed SST v5 data are available from NOAA National Centers for Environmental Information. IERS EOP C04 20 LOD data are available from the IERS Earth Orientation Centre (https://www.iers.org/IERS/EN/DataProducts/EarthOrientationData/eop.html). Derived indices, composite data, and the analysis code used to produce the figures will be deposited in a public repository upon acceptance; in the interim they are available from the corresponding author upon reasonable request.

## APPENDIX A

### Full 12-index regional definitions

SLP and wind anomalies are computed relative to the 1963–2023 monthly climatology. LOD anomalies are computed relative to the 1963–2022 monthly climatology. Indices marked with a star (★ ) are important in the main text.

**TABLE A1.** *Full 12-index regional definitions.*

| Index | Latitude | Longitude | Variable | Role |
|---|---|---|---|---|
| SLP_R1 | 20–30°N | 180–150°W | SLPa | NH EPSH, western sector |
| SLP_R2 ★ | 20–30°N | 150–120°W | SLPa | NH EPSH, central |
| SLP_R3 | 20–30°N | 120–90°W | SLPa | NH EPSH, eastern sector |
| SLP_R4 | 20–30°S | 180–150°W | SLPa | SH EPSH, western sector |
| SLP_R5 ★ | 20–30°S | 150–120°W | SLPa | SH EPSH, central |
| SLP_R6 | 20–30°S | 120–90°W | SLPa | SH EPSH, eastern sector |
| U_R1 ★ | 2.5°S–2.5°N | 160–180°E | Ua | Central equatorial wind |
| U_R2 | 2.5°S–2.5°N | 170–150°W | Ua | Central-east equatorial |
| U_R3 | 2.5°S–2.5°N | 140–110°W | Ua | Eastern equatorial |
| NINO3 ★ | 5°S–5°N | 150–90°W | SSTa | ENSO index |
| SST_R1 ★ | 2.9°S–2.9°N | 180–150°W | SSTa | Central Pacific SST |
| SST_R2 | 2.9°S–2.9°N | 150–120°W | SSTa | Central-east Pacific SST |
| SST_R3 | 2.9°S–2.9°N | 120–90°W | SSTa | Eastern Pacific SST |

# Supplemental Material

*Annually Modulated Pacific Subtropical Highs Pace the ENSO Cycle: A Hierarchical Framework from Earth Rotation to Atmosphere–Ocean Dynamics*

This document collects the detailed calculations, statistical robustness checks, and supporting figures for the LOD–EPSH–ENSO causal chain summarized in the main text. Figures and tables here use an S prefix (Fig. S1, S2, …; Table S1, S2, …), independent of the main text's numbering, and are pointed to from the relevant places in the main text.

**1. Five-band spectral decomposition of length of day**

FIG. S1. Five-band spectral decomposition of the de-tided LOD anomaly series, 1963–2022. (a) Full de-tided LOD anomaly ($\sigma$ = 0.839 ms), from which the six astronomical tidal components have been removed. The long-term decline reflects secular tidal braking and core–mantle coupling. (b) 0–0.5 yr high-pass component (variance = 1.1%, $\sigma$ = 0.086 ms). (c) 0.5–1 yr sub-annual component (0.6%, $\sigma$ = 0.064 ms, dominant period about eight months), the band carrying the upstream ENSO precursor signal. (d) 1–1.5 yr transition band (0.3%, $\sigma$ = 0.043 ms). (e) 1.5–2 yr quasi-biennial band (0.3%, $\sigma$ = 0.042 ms). (f) 2–10 yr ENSO-band interannual component (3.2%, $\sigma$ = 0.151 ms), carrying the well-established ENSO → LOD response. Red and blue vertical lines mark El Niño and La Niña event years, respectively. The secular component (> 10 yr, ~91% of variance) is not shown.

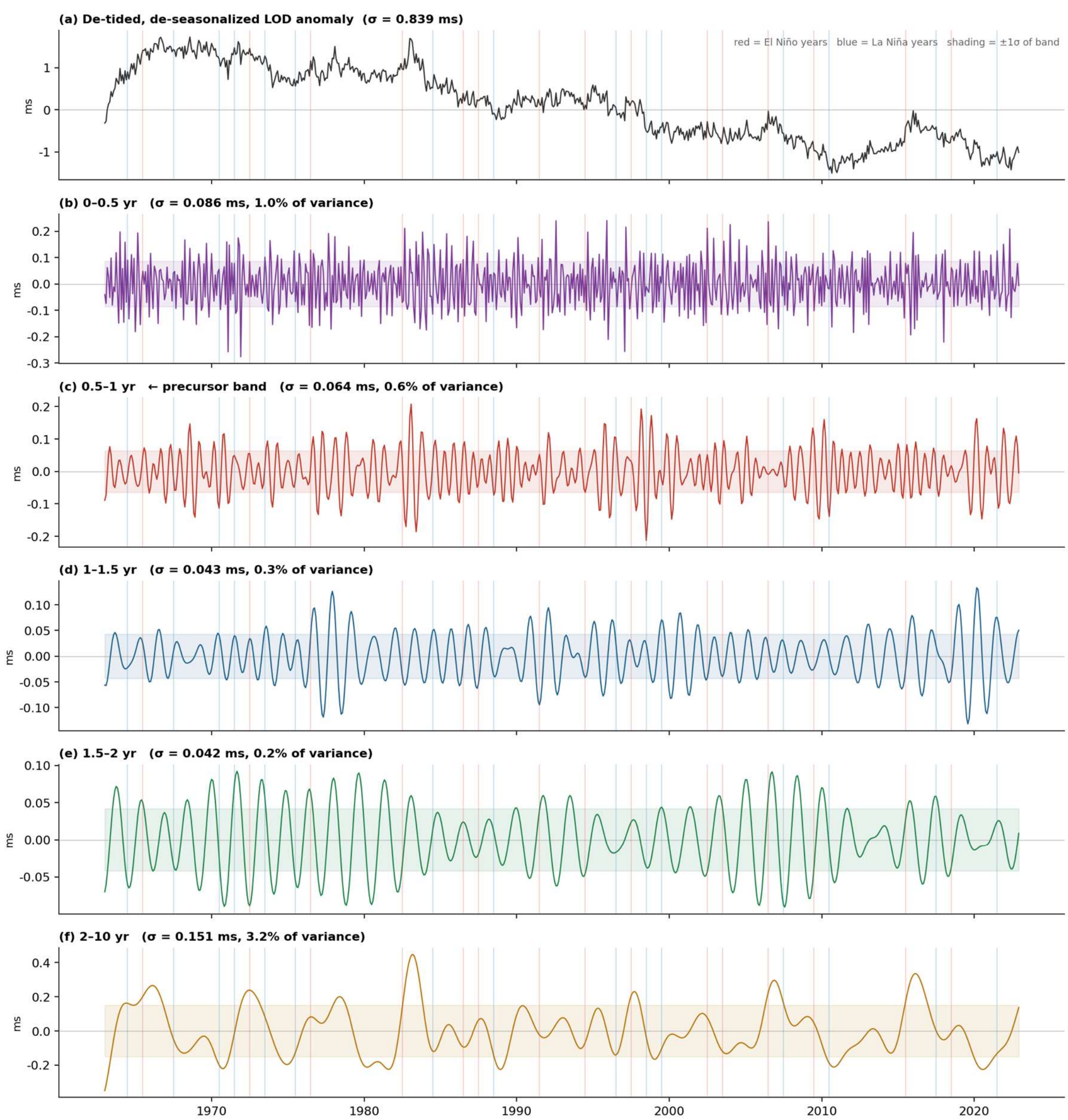

## 2. LOD filter specifications and seasonal bias test

All filters are fourth-order zero-phase Butterworth implemented as forward–backward filtfilt (Butterworth 1930; Oppenheim and Schafer 2009). Endpoint ringing (Gibbs phenomenon) is suppressed by AR(15) autoregressive extension of 72 months on each side (Press et al. 2007). The lower bound of band (i) is set by the Nyquist constraint of monthly data (Nyquist 1928; Shannon 1949). An independent approach to LOD parameter extraction consistent with the present decomposition is described in Xu et al. (2022).

To verify that the LOD(0.5–1 yr) filtered series carries no spurious seasonal cycle, we compute the monthly climatological mean and apply a one-sample t-test against zero. All 12 monthly means are statistically indistinguishable from zero (all $|t| < 0.29$; all $P > 0.78$). The largest monthly mean is 0.003 ms (November), representing 4.1% of the band's standard deviation (0.064 ms). The bandpass filter effectively removes the astronomical tidal signals pre-extracted in the de-tiding step. Table B1 summarizes filter specifications and variance contributions across all five bands. The treatment of the seasonal harmonics deserves emphasis: the annual and semi-annual harmonics are removed deterministically in the de-tiding step, precisely to guard against contamination of the 0.5 to 1 yr band by the seasonal AAM cycle (Egger et al. 2007). The dominant period of the retained band is about eight months rather than the band midpoint: the median crest-to-crest spacing is 8 months over the 89 complete cycles of 1963 to 2022 (interquartile range 7 to 9 months), the power-weighted mean period is 8.4 months, and the crests fall in all twelve calendar months with near-uniform frequency, confirming that the oscillation is not phase-locked to the calendar. Repeating the entire chain analysis without removing these harmonics changes the headline correlations only marginally (the Y−1 December ridge correlation moves from +0.412 to +0.415 and the Y0 May northern-ridge correlation is unchanged at +0.380, the two A2 versions correlating at 0.999); the precursor therefore does not rest on the seasonal treatment.

TABLE S1. LOD bandpass filter specifications and variance contributions, 1963–2022.

| Band | Period range | Variance (%) | σ (ms) |
|---|---|---|---|
| (i) High-pass | 0–0.5 yr | 1.1 | 0.086 |
| (ii) Sub-annual | 0.5–1 yr | 0.6 | 0.064 |
| (iii) Transition | 1–1.5 yr | 0.3 | 0.043 |
| (iv) Quasi-biennial | 1.5–2 yr | 0.3 | 0.042 |
| (v) ENSO-band | 2–10 yr | 3.2 | 0.151 |
| Secular (residual) | >10 yr | ~91 | - |
| Total (de-tided) | All | 100 | 0.839 |

**3. The causal chain, link by link**

FIG. S2. The causal chain, link by link (n = 59 years, 1964–2022). Red: El Niño years; blue: La Niña years; gray: neutral years. Solid lines are ordinary least squares fits; significance is assessed by phase-randomized Monte Carlo (effective sample size ≈ 50; 95% threshold $|r| \approx 0.27$). (a) A2 versus SLP_R2+R5a in Y−1 December, the stage at which the eastern-Pacific ridges respond as a coherent system ($r = +0.41$, $P = 0.0012$). (b) A2 versus SLP_R2a in Y0 May, the stage at which the northern ridge dominates ($r = +0.38$, $P = 0.0032$). (c) SLP_R2+R5a (Y0 May) versus U_R1a (Y0 June): the ridge anomaly sets the central equatorial zonal wind ($r = -0.47$, $P = 0.0002$). (d) U_R1a (Y0 June) versus SST3a (Y0 SON): the wind anomaly sets the ocean through the Bjerknes

feedback (r = +0.63, P < 0.0001).

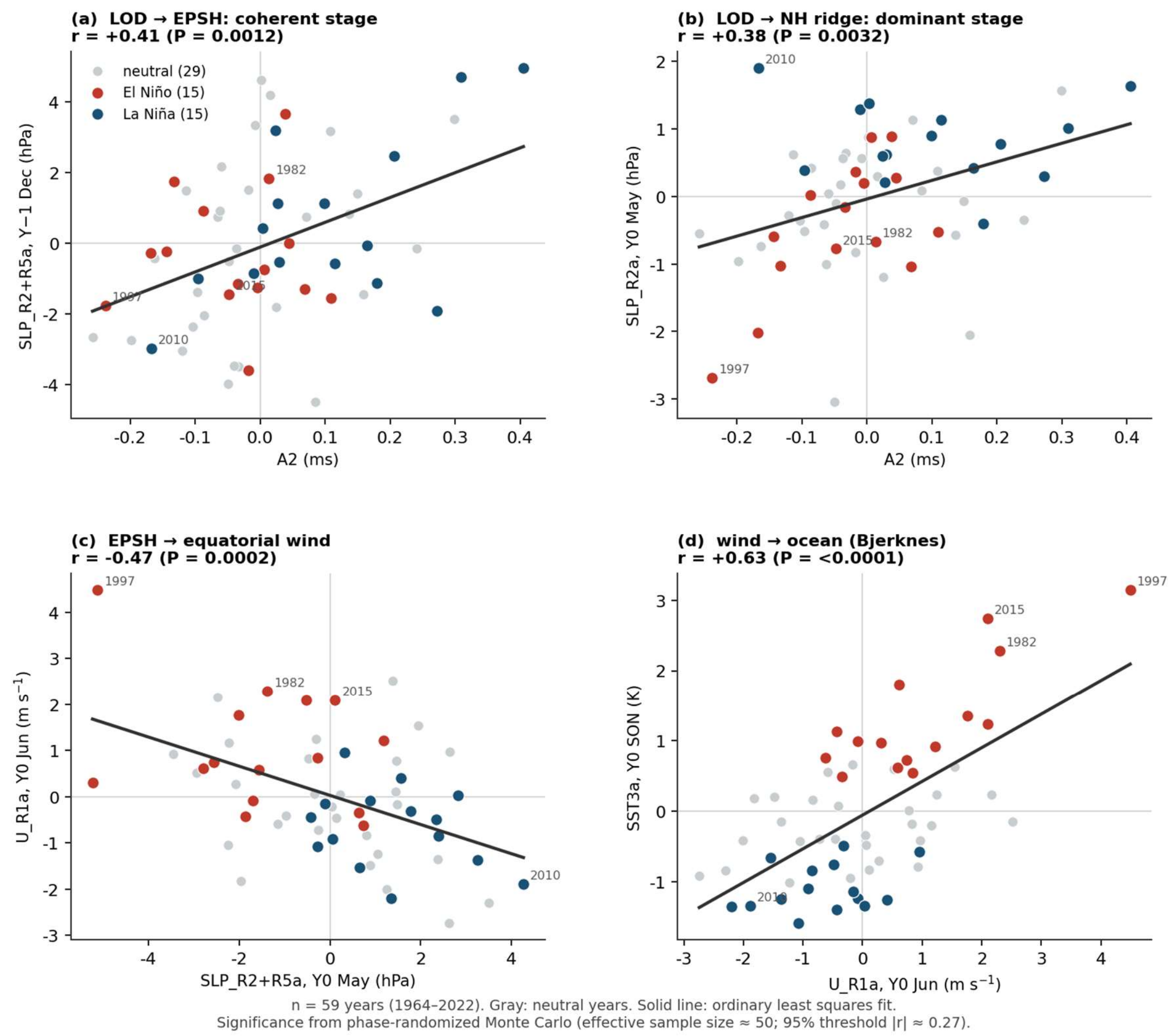

## 4. Full mediation table

TABLE S2. Mediation by partial correlation, n = 59 years (1964–2022). The first row gives the total effect c (A2 → SST3a with nothing held fixed); the following rows give the direct residual c′ when each candidate mediator is held fixed; the final column gives path b, the mediator-to-ocean partial correlation with A2 controlled. A significant total effect together with a non-significant direct residual under every choice of mediator is the signature of complete mediation. Bootstrap product-of-coefficients results (5,000 resamples) corroborate every decomposition and are summarized in appendix C.

| Control variable | A2 → SST3a (SON) | A2 → SST3a (DJF) | EPSH → SST3a \| A2 | Verdict |
|---|---|---|---|---|
| none (total effect c) | − 0.31 (P = 0.017) | − 0.33 (P = 0.010) | − 0.53 | significant |
| SLP_R2a (Y0 | − 0.14 (P = | − 0.16 (P = | — | c′ not |

| May) | 0.31) | 0.22) | | significant |
|---|---|---|---|---|
| SLP_R2+R5a (Y0 May) | − 0.15 (P = 0.25) | − 0.18 (P = 0.18) | − 0.47 (P = 0.0002) | c′ not significant |
| U_R1a (Y0 June) | − 0.17 (P = 0.21) | − 0.20 (P = 0.14) | — | c′ not significant |
| Path a: A2 → SLP_R2+R5a (May) | +0.36 (P = 0.005) | | | mediator engaged |
| Path b: SLP_R2+R5a → U_R1a \| A2 | − 0.41 (P = 0.0015) | | | mediator transmits |

**5. Five independent lines of evidence for the causal chain**

The evidence for the LOD–EPSH–ENSO causal chain rests on five independent and mutually reinforcing analyses, each addressing different potential confounds.

(i) Spectral composite analysis (Fig. 6): near-mirror-image 180°-out-of-phase polarity in four of five LOD spectral bands persists across 30 diverse ENSO events spanning weak and strong, EP and CP, and PDO-positive and PDO-negative conditions. This systematic pattern cannot be attributed to the sampling of a few high-impact events.

(ii) Full-sample lead–lag correlation (Fig. 7): A2 versus SLP_R2+R5a reaches $r = +0.41$ ($P = 0.0012$, $n = 59$) in Y−1 December and $r = +0.36$ ($P = 0.005$) in Y0 May, with the temporal evolution (SH EPSH response in austral late spring, NH EPSH response in boreal late spring, U_R1a significant in June, SSTa peaking in autumn) consistent with a forward-propagating causal chain, not a simultaneous common-forcing response.

(iii) Mediation analysis (Table S2; Fig. 9): the total effect is significant (A2 → SST3a: $r = -0.31$ in SON, −0.33 in DJF), the mediator is engaged and transmits (path a = +0.36, $P = 0.005$; path b = −0.41 with A2 controlled, $P = 0.0015$), and the direct residual path collapses to between −0.14 and −0.20 ($P$ = 0.14 to 0.31) under every choice of mediator, all by partial correlation with no resampling. Combined with the strict temporal ordering established in (ii), this pattern supports a forward-propagating chain and is difficult to reconcile with reverse causality.

(iv) Bootstrap corroboration: a product-of-coefficients analysis (5,000 resamples; Preacher and Hayes 2008) reproduces the same verdict in every configuration, with indirect effects whose 95% confidence intervals exclude zero, a mediated share of approximately one half rising to 55–60% through the northern (R2) channel alone, and SLP_R2+R5a mediating 61% of the A2–U_R1a association (bootstrap $P = 0.016$), confirming the EPSH as the atmospheric intermediary.

(v) The 1997/98 case study (Figs. 4b, 5b, and S4): the correct temporal ordering; strong positive LOD(0.5–1 yr) phase (summer 1996), coordinated EPSH weakening (late 1996–early 1997), westerly wind bursts (December 1996–March 1997), SST3a peak (December 1997); is documented at 2–3 month resolution in a well-studied, independently verified event.

Furthermore, the significance of A2 across 59 years spanning weak, strong, EP, CP, and PDO-modulated events provides stronger evidence for a robust physical mechanism than the same correlation derived from a homogeneous subset of strong events alone. The diversity of the 59-year sample is itself a source of statistical strength.

**6. Sensitivity to outlier removal, forking paths, and resampling stability**

A natural concern is whether the chain is driven by a small number of extreme ENSO events. We recomputed all main correlations with the three most extreme El Niño events removed (1982/83, 1997/98, 2015/16; n = 56). The chain survives with graded strength: A2 vs. SLP_R2a (Y0 May): +0.38 → +0.32 (P = 0.016); A2 vs. SLP_R2+R5a (Y0 May): +0.36 → +0.31 (P = 0.022); A2 vs. SLP_R2+R5a (Y−1 Dec): +0.41 → +0.40 (P = 0.002), the most robust link in the profile; SLP_R2+R5a vs. U_R1a: −0.47 → −0.38 (P = 0.004); U_R1a vs. SST3a: +0.63 → +0.46 (P = 0.0004). The direct A2 vs. NINO3 weakens to marginal in SON (−0.25, P = 0.062) but remains significant in DJF (−0.29, P = 0.031). The A2 vs. SLP_R5a (Y0 May) link does not survive removal (+0.20, P = 0.15), whereas the R5 response at Y−1 December strengthens (+0.35 → +0.38, P = 0.004), confirming that the SH channel operates in austral late spring rather than in Y0 May. The chain thus does not rest on a few extreme events, though individual links range from highly robust (Y−1 Dec EPSH; U → SST) to moderately robust (Y0 May; DJF NINO3).

TABLE S3. Sensitivity of chain correlations to removal of the three strongest El Niño events.

| Correlation | Full sample (n = 59) | Without 1982, 1997, 2015 (n = 56) | P after removal | Verdict |
|---|---|---|---|---|
| A2 vs. SLP_R2+R5a (Y- 1 Dec) | +0.412 | +0.402 | 0.002 | highly robust |
| A2 vs. SLP_R2a (Y0 May) | +0.380 | +0.321 | 0.016 | robust |
| A2 vs. SLP_R2+R5a (Y0 May) | +0.359 | +0.306 | 0.022 | robust |
| A2 vs. SLP_R5a | +0.233 | +0.197 | 0.146 | not robust (SH |

| (Y0 May) | | | | acts earlier) |
|---|---|---|---|---|
| A2 vs. NINO3 (Y0 DJF) | - 0.332 | - 0.288 | 0.031 | moderately robust |
| SLP_R2+R5a (May) vs. U_R1a (Jun) | - 0.470 | - 0.381 | 0.004 | robust |
| U_R1a (Jun) vs. SST3a (SON) | +0.625 | +0.456 | <0.001 | highly robust |

The March and July sampling months were suggested by the same 30-event composite that motivates A2, so the nominal p-values of the fixed window are potentially optimistic, the selection effect statisticians call the garden of forking paths. Three checks address this directly. First, the window family is smooth rather than peaked: across the nine one-month variants (start February to April, end June to August) the correlation with the following DJF NINO3 ranges from −0.24 to −0.35 and with the Y−1 December ridge index from +0.32 to +0.41, and the fixed window is not even the maximum at the SST endpoint. Second, a selection-aware Monte Carlo, in which each of 5,000 phase-randomized surrogates is allowed the same nine-window search and the maximum |r| is retained, gives P = 0.008 for the Y−1 December EPSH response (fixed-window P = 0.002) and P = 0.07 for the DJF NINO3 endpoint (fixed-window P = 0.034). Charged for the search, the SST endpoint becomes marginal while the EPSH response remains clearly significant, which is precisely the ordering the chain architecture asserts: the load-bearing link is LOD to EPSH, the ocean is reached through the EPSH to wind to SST links, which involve no window selection, and the direct A2 to SST3a correlation is corroborative rather than structural.

Third, the relationship is stable under resampling of years. Splitting the record in half gives r = +0.30 (1964 to 1992) and +0.49 (1993 to 2022) for the December EPSH link, and −0.21 and −0.43 for the NINO3 endpoint. Leaving out each of the 30 events one at a time keeps the EPSH correlation within +0.34 to +0.46 and the NINO3 correlation within −0.27 to −0.36, with no sign change in 60 recomputations. Deleting each decade in turn keeps the EPSH link within +0.34 to +0.47 and the NINO3 link within −0.22 to −0.40. By every cut the December ridge response behaves as the most robust link in the profile, consistent with Table S3.

**7. Thirty-six-month composite extending into the Y+1 decay phase**

The 24-month composite in Fig. 7 focuses on the precursor year Y−1 and the ENSO year Y0, where formal significance testing with Welch's t-test is applied. Extending the window to include the decay year Y+1 provides a visual display of the full temporal succession from LOD as preconditioner (Y−1), through MSIGMA as terminator (Y0 autumn onward), to LOD as response (Y+1). The extended composite is presented here rather than in the main text because its LOD panel (Fig. E1a) overlaps with the information already contained in Fig. S1c (24-month spectral composites, 0.5–1 yr band) and Fig. 7a (24-month five-variable composite, LOD panel); the novel content of Fig. E1 is therefore restricted to the Y+1 window. No formal significance testing is applied to the Y+1 portion, because the LOD signal in that window is partly driven by the ENSO event itself through the ENSO → AAM → LOD feedback.

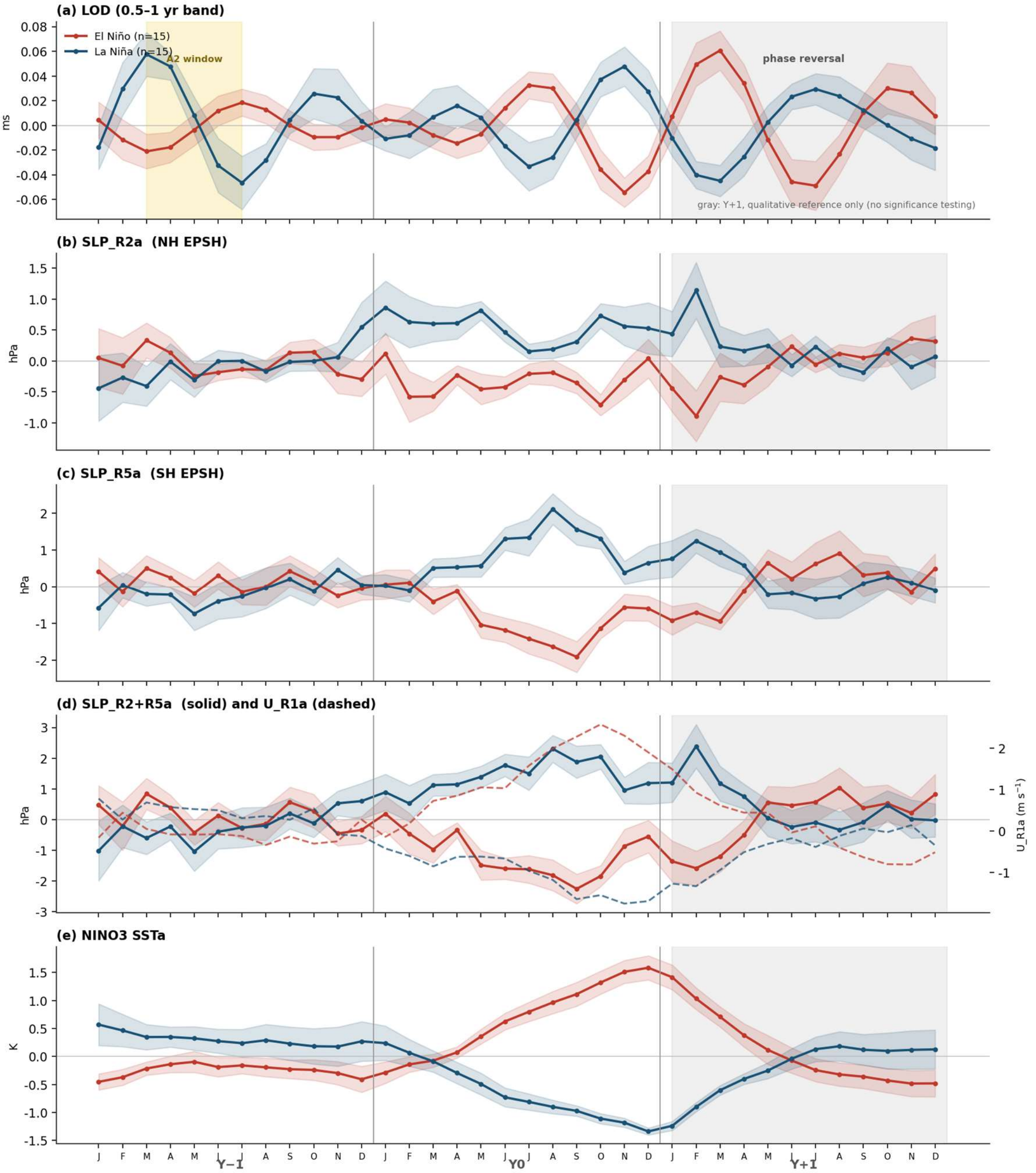

FIG. S3. Thirty-six-month composites (Y−1 January to Y+1 December) of the fifteen El Niño and fifteen La Niña events, with the same five-variable layout as Fig. 8 extended to include Y+1 (months 25 to 36). Shading around each curve is ±1 standard error of the mean. Gray shading over Y+1 indicates qualitative reference only; no formal significance testing is applied there, because the length-of-day signal in Y+1 is partly driven by the ENSO event itself through the ENSO to AAM to LOD response. (a) LOD (0.5–1 yr band), with the A2 sampling window shaded. The spring-to-summer swing reverses sign between the precursor year and the year after the event: averaged over the warm events the swing is −0.04 ms in Y−1 and +0.11 ms in Y+1, and averaged over the cold events it is +0.10 ms in Y−1 and −0.07 ms in Y+1. A rising length of day before a warm event is thus followed, two years later, by a falling one, and conversely for cold events. We read this reversal as the geodetic expression of the Mirror-SIGMA termination mechanism, which resets the phase of the oscillation as the event decays. (b, c) SLP_R2a and SLP_R5a: the ridge anomalies peak in Y0 and decay through Y+1. (d) SLP_R2+R5a (solid) with U_R1a (dashed, right axis), the wind anomaly reversing during ENSO decay. (e) NINO3 SSTa, peaking in Y0 boreal autumn and decaying through Y+1. The opposite sign of the Y−1 precursor and the Y+1 response is why earlier undifferentiated LOD-ENSO correlation analyses recovered weak net signals (Zheng et al. 2003): the two contributions partially cancel when all lags are combined without separating the temporal windows.

**8. Case-level LOD–EPSH verification for the 1997/98 super El Niño**

The 30-event composite (Fig. 7) and the full-sample mediation analysis (Figs. 7–9, Table S2) demonstrate that the LOD → EPSH → wind → SST causal chain is statistically robust. A complementary test is whether the same sequence can be read directly from a single, well-documented event. The 1997/98 super El Niño provides such a case. Figure E2 displays the five-variable evolution from January 1995 through October 1999: LOD (0.5–1 yr bandpass) at the top, followed by SLP_R2a, SLP_R5a, U_R1a, and NINO3 SSTa. Arranged vertically in the order of the causal chain, the figure makes the sequential temporal phasing visually explicit. The LOD peak in June 1996, with the A2 window closing the following month, precedes the December weakening of the northern ridge by about five months and the May weakening by about eleven, two of the three stages of the composite analysis. The WWB cluster (December 1996–March 1997) occurs during the early phase of EPSH weakening, illustrating how Tier 2b intraseasonal triggers act on a background state preconditioned by Tier 2a (LOD/AAM) and Tier 3 (EPSH). The SST peak follows the closure of the A2 window by seventeen months, so length-of-day observations available by mid-1996 already carried the phase information, well ahead of the boreal spring 1997 predictability barrier. This single-event view complements the statistical evidence in Figs. S1–S2, 7–9 and shows that the relationships are not compositing artifacts.

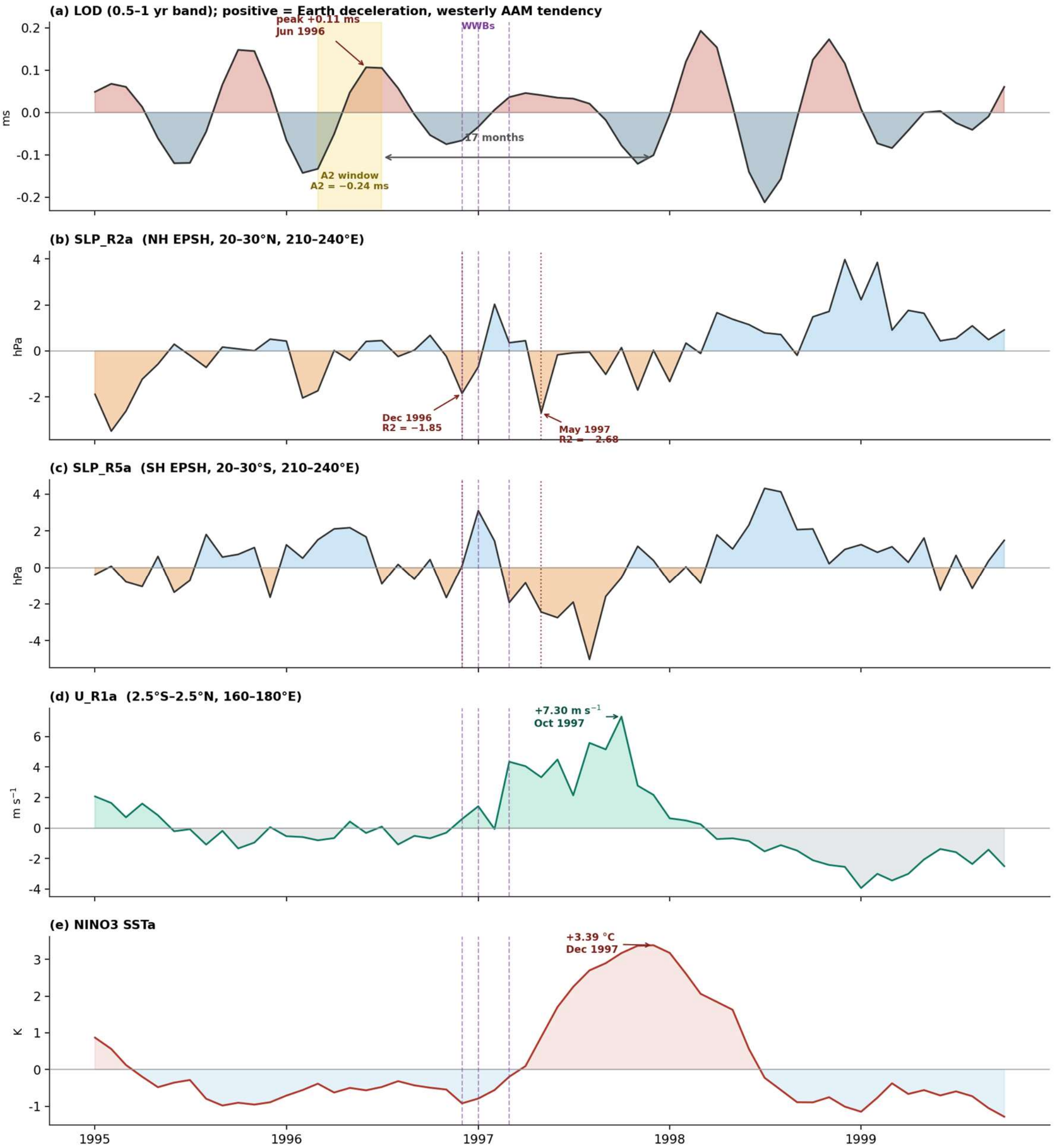

FIG. S4. Case-level illustration of the LOD to EPSH to U_R1a to SST3a sequence for the 1997/98 El Niño, January 1995 through October 1999. (a) LOD (0.5–1 yr band; ms), positive values indicating a lengthening day, that is, Earth deceleration and an atmosphere gaining westerly angular momentum. The oscillation reached a positive peak of +0.11 ms in June 1996 and was still +0.10 ms in July, so that with the March value at −0.13 ms the index is A2 = −0.24 ms, the second

most negative of the fifty-nine years. Red and blue shading distinguish positive from negative phases; the yellow band marks the A2 sampling window. The three purple dashed lines mark the westerly wind bursts of December 1996, January 1997 and March 1997 (McPhaden 1999). (b) SLP_R2a and (c) SLP_R5a (hPa); orange shading marks ridge weakening, which favors El Niño, and blue shading ridge strengthening. Two of the three stages identified in the composite analysis are clearly expressed in this case: the northern ridge is strongly weakened in December 1996 (SLP_R2a = −1.85 hPa, and SLP_R2+R5a = −1.75 hPa) and again in the following May (SLP_R2a = −2.68 hPa, SLP_R2+R5a = −5.12 hPa). The August stage is not expressed in this particular year (SLP_R5a = +0.17 hPa in August 1996), and individual months between the stages fluctuate in sign, as single realizations of a modest correlation must; the composite of Fig. 9, not any one case, is the basis of the argument. (d) U_R1a (m $s^{-1}$), the peak westerly anomaly of +7.30 m $s^{-1}$ in October 1997. (e) NINO3 SSTa (°C), peaking at +3.39°C in December 1997. The A2 index is complete at the end of July 1996, seventeen months before that peak.

**9. Detailed derivations supporting the discussion in the main text**

*9a. Angular-momentum sign verification (main text section 7b)*

Each sign in the chain follows: A2 is negative when length of day rises from March to July of the precursor year, so a negative A2 marks an atmosphere gaining westerly angular momentum through that spring; and negative A2 is indeed followed by negative sea level pressure anomalies at both ridges ($r = -0.37$ at R5 in August, −0.41 at R2+R5 in December, −0.38 at R2 in the following May, when the correlations are recast against the rise in length of day), by a westerly anomaly in the equatorial zonal wind, and by a warm event ($r = -0.33$ with the following DJF NINO3, here in A2's own sign convention rather than recast). Not one sign in the sequence has to be assumed; each is measured, and each is the sign that conservation of angular momentum requires. Zonal-mean angular-momentum anomalies are observed to propagate poleward, accompanied by anomalous eddy flux convergence, on both intraseasonal and interannual timescales (Dickey et al. 1992; Weickmann et al. 1997; Egger et al. 2007), which is the behavior the correlation fields of Fig. 7 display between August and the following May.

*9b. The annual-harmonic diagnostic and band orthogonality (main text sections 5b and 8)*

The classical ENSO-driven response of Earth rotation lives in the 2 to 10 yr band, where band-passed length of day lags NINO3 by about a month at $r = 0.40$; in the 0.5 to 1 yr band used throughout this paper the same cross-correlation never exceeds 0.05 at any lag. ENSO does imprint on Earth rotation through the annual cycle: fitting the annual harmonic year by year, its amplitude covaries with the following winter NINO3 at $r = -0.31$ ($P = 0.015$, $n = 60$), and that harmonic, with a mean amplitude of 0.390 ms, is six times the standard deviation of the sub-annual band itself and sits directly on its edge, inflating the 0.5–1 yr band threefold if left in place. We remove the annual and semi-annual harmonics deterministically for this reason. The orthogonality does not depend on that step: with the harmonics retained the maximum cross-correlation of the band with NINO3 is 0.015, and with them removed it is 0.050; either way the band is deaf to ENSO amplitude. The preceding event does organize the sub-annual band through phase rather than amplitude: A2 correlates with the previous winter's NINO3 at +0.64 while sharing almost no monthly amplitude coherence with it, consistent with the termination of an event resetting the oscillation's phase without feeding energy into it. That the downstream chain is the re-phased oscillation's own doing, and not the memory of the last event, follows from holding the previous winter's NINO3 fixed: the December ridge response is entirely undiminished (partial $r = +0.41$), the May northern-ridge response survives (partial $r = +0.31$), and the equatorial-wind response strengthens (partial $r = -0.34$).

*9c. Event-level rescue statistics behind Fig. 8 (main text section 7c)*

With the fixed window, 21 of the 30 classified events are correctly signed. Seven of the nine misses are recovered by sliding the whole four-month window a single month, always to February through June or April through August and never further, the displacement expected of an eight-month oscillation unlocked from the calendar. The clearest case is the 1982/83 event: the precursor-year crest fell in May 1981, midway between the sampling months, leaving A2 within 0.2 standard deviations of zero before one of the strongest El Niño events on record; the index issued silence rather than a wrong call. Two cold events, 1973 and 2010, resist any admissible shift and are reported as failures. The recovery is retrospective and is not a forecast rule, but it shows that most of the miss rate reflects the timing of the sampling rather than the absence of a signal.